\documentclass[]{aa}

\usepackage{multirow}
\usepackage{mwe}
\usepackage[utf8]{inputenc}
\usepackage[varg]{txfonts}
\usepackage[T1]{fontenc}
\usepackage{graphicx}
\usepackage{amsmath}
\usepackage{amssymb} 
\usepackage{hyperref}
\usepackage{natbib}
\usepackage{hyperref}
\usepackage{enumerate}
\usepackage{adjustbox}
\usepackage{fancyref}

\usepackage[table]{xcolor}
\definecolor{gray75}{gray}{0.75}
\definecolor{dkgreen}{rgb}{0,0.6,0}
\definecolor{gray}{rgb}{0.5,0.5,0.5}
\definecolor{citationcolor}{rgb}{0,0.5,0}
\definecolor{linkcolor}{rgb}{0.5,0,0}

\hypersetup{colorlinks=true, breaklinks=true, linkcolor=linkcolor, menucolor=linkcolor, urlcolor=linkcolor,citecolor=citationcolor} 
\usepackage{longtable}
\usepackage[para]{threeparttable}
\usepackage[para]{threeparttablex}
\addtocounter{footnote}{1}
\usepackage{mathrsfs}
\usepackage{paralist}
\usepackage{url}
\usepackage{siunitx}
\usepackage{setspace}
\usepackage{multicol}
\usepackage[bf]{caption}
\usepackage{booktabs,makecell} 
\usepackage[]{float}
\usepackage{pdfpages}

\usepackage{graphicx}
\usepackage[varg]{txfonts}
\bibpunct{(}{)}{;}{a}{}{,}

\title{Probing the disk-jet coupling in M\,87}
\author{{Ainara Saiz-P\'{e}rez \inst{\ref{affil:wuerzburg}}\fnmsep\thanks{ainara.saiz-perez@uni-wuerzburg.de}}
        \and {Christian M. Fromm \inst{\ref{affil:wuerzburg},\ref{affil:frankfurt},\ref{affil:mpifr}}\fnmsep\thanks{christian.fromm@uni-wuerzburg.de}}
        \and Yosuke Mizuno \inst{\ref{affil:tsung},\ref{affil:jiaotong},\ref{affil:keylab},\ref{affil:frankfurt}}
        \and Matthias Kadler \inst{\ref{affil:wuerzburg}}
        \and Karl Mannheim \inst{\ref{affil:wuerzburg}}
        \and Ziri Younsi\inst{\ref{affil:mullard}}
}
\institute{%
        Institut f\"ur Theoretische Physik und Astrophysik, Universit\"at W\"urzburg, Emil-Fischer-Str. 31, D-97074 W\"urzburg, Germany\label{affil:wuerzburg}
        \and%
        Institut f\"ur Theoretische Physik, Goethe Universit\"at, Max-von-Laue-Str. 1, D-60438 Frankfurt, Germany \label{affil:frankfurt}
        \and%
        Max-Planck-Institut f\"ur Radioastronomie, Auf dem H\"ugel 69, D-53121 Bonn, Germany\label{affil:mpifr} 
        \and 
        Tsung-Dao Lee Institute, Shanghai Jiao Tong University, 1 Lisuo Road, Shanghai 201210, People’s Republic of China \label{affil:tsung}
        \and
        School of Physics \& Astronomy, Shanghai Jiao Tong University, 800 Dongchuan Road, Shanghai 200240, People’s Republic of China \label{affil:jiaotong}
        \and
        Key Laboratory for Particle Physics, Astrophysics and Cosmology, Shanghai Key Laboratory for Particle Physics and Cosmology, Shanghai Jiao-Tong University, 800 Dongchuan Road, Shanghai 200240, People’s Republic of China\label{affil:keylab}
        \and
        Mullard Space Science Laboratory, University College London, Holmbury St. Mary, Dorking, Surrey RH5 6NT, UK\label{affil:mullard}
}%

\begin{document}

        \abstract{
                \textit{Context.} Recent GMVA observations of M\,87 at event horizon scales revealed a ring-like structure that is $\sim50\%$ larger at 86\,GHz than the ring observed by the Event Horizon Telescope at 230\,GHz.  \newline 
                \textit{Aims.} We studied a possible origin of the increased ring size at 86\,GHz and the role the nonthermal electron population plays in the observed event horizon scales. \newline
                \textit{Methods.} We carried out 3D general relativistic magnetohydrodynamic simulations followed by radiative transfer calculations. We incorporated synchrotron emission from both thermal and nonthermal electrons into the calculations. To better compare our results to observations, we generated synthetic interferometric data adjusted to the properties of the observing arrays. We fit geometrical models to these data in Fourier space through Bayesian analysis to monitor the variable ring size and width over the simulated time span. \newline
                \textit{Results.} We find that the 86\,GHz ring is always larger than the 230\,GHz ring, which can be explained by the increased synchrotron self-absorption at 86\,GHz and the mixed emission from both the accretion disk and the jet footpoints, as well as flux arcs ejected from a magnetized disk. We find agreement with the observations, particularly within the error range of the observational value of $M/D$ for M\,87. \newline
                \textit{Conclusions.} We show that state-of-the art 3D general relativistic magnetohydrodynamic simulations 
                combined with thermal and nonthermal emitting particles can explain the observed frequency-dependent ring size in M\,87. Importantly, we find that MAD events triggered in the accretion disk can significantly increase the lower-frequency ring sizes.}
        
        \maketitle
        
        \section{Introduction}
        General relativity predicts that event horizons exist generically in black holes, regardless of geometry \citep{penrose1965}. \citet{LyndenBell1969} proposed that supermassive black holes (SMBHs) could explain some of the observables at the center of galaxies, an idea later supported by measurements of stellar and gas dynamics in nearby galaxies \citep{Kormendy1995}. Most galaxies are now believed to contain an accreting SMBH capable of launching a relativistic jet that is observable at subparsec to megaparsec scales \citep{blandford2019}. These objects, known as active galactic nuclei (AGNs), are able to outshine their host galaxy.
        
        Due to its proximity and bright jet, the elliptical galaxy M\,87 provides an ideal laboratory for studying accretion and jet launching around SMBHs. M\,87 appears on the sky as a bright, compact radio source, a characteristic feature of low-luminosity AGNs (\citealt{nagar2005}). \citet{dimatteo2003} furthermore estimated that the accretion of matter onto the central SMBH of M\,87 is radiatively inefficient, and low-luminosity AGNs are often modeled through advection-dominated accretion flow models \citep{narayan1995,yuan2014}. These systems describe accretion through an optically thin, geometrically thick torus, the ``disk,'' where most of the viscously dissipated energy is advected rather than radiated away, resulting in a luminosity lower than expected. \citet{yuan2002} first proposed a disk-jet setup to explain the spectrum of a prototypical low-luminosity AGN at low frequencies.
        
        The jet of M\,87 was first observed at optical frequencies by \citet{curtis1918} and has since been extensively studied at wavelengths from the radio to $\gamma$-rays \citep[][]{abramowski2012,MAGIC2020}. Very long-baseline interferometry (VLBI) observations at millimeter wavelengths of the innermost core regions of M\,87, $7-100\,R_\mathrm{S}$ (where $R_\mathrm{S}=2GM/c^2$ is defined by the speed of light, $c$, the gravitational constant, $G,$ and the mass of the black hole, $M$), show a limb-brightened jet with a wide opening angle and parabolic collimation, supporting the picture of a jet launched magnetically around a black hole \citep{reid1989,junor1999,hada2016,mertens2016,kim2018,walker2018}. The apparent position of the radio core shifts upstream with increasing frequency, indicating that the core is becoming optically thin to synchrotron emission. \citet{hada2011} studied the core shift of M\,87 and determined that the black hole overlaps with the observed radio core at frequencies $\geq$~43GHz, as the peak brightness should coincide with the position of the central engine \citep{blandford1979}. 
        
        Millimeter and submillimeter VLBI allows us to directly image the event horizon scales of AGNs, where the central SMBHs are expected to cast a shadow \citep{falcke2002}. Only Sgr\,A$^\star$ and M\,87 have an angular size large enough to be resolved by our current submillimeter VLBI technology. The event horizon region of M\,87 was first resolved at 230~GHz by the Event Horizon Telescope (EHT), which found an asymmetric ring-like structure with a diameter of $42\pm3\,\mu$as and a central depression. This morphology is consistent with the theoretical models of accretion onto a rotating black hole and remains present in more recent observations \citep{EHTa2019,EHTshadowI2024}. Observations of Sgr\,A$^\star$ by the EHT provided further evidence in favor of the SMBH picture \citep{EHTa2022}, but unlike Sgr\,A$^\star$, M\,87 hosts a powerful, extended jet. Due to this, M\,87 remains the only source for which we can hope to simultaneously resolve the extended jet and the event horizon structures. This was achieved by \citet{lu2023} with 86~GHz observations by the Global Millimeter VLBI Array (GMVA) together with the Atacama Large Millimeter/submillimeter Array (ALMA) and the Greenland Telescope (GLT). They found a ring-like structure with a diameter ($d$) of $64^{+4}_{-8}\,\mu$as, which is $\sim$50\% larger than that at 230~GHz. More recently, \citet{Kim2025} applied two novel imaging algorithms to the 86~GHz data and found lower, but compatible, diameter values of $60.9\pm2.2\,\mu$as and $61.0\,\mu$as.
        
        General relativistic magnetohydrodynamic (GRMHD) simulations can model accretion onto a black hole and the subsequent jet launching. Coupled with general relativistic radiative transfer (GRRT) calculations that account for synchrotron emission, we can obtain synthetic images that can be compared to VLBI observations. The EHT Collaboration extensively studied which disk-jet setups could explain the observed intensity and polarization images at 230~GHz \citep{EHTe2019,EHTh2021}. They find that only models with strongly magnetized disks, known as magnetically arrested disks (MADs), remain viable, but that several values of the black hole spin ($a_*$) can explain the observed polarization patterns. 
        
        \citet{lu2023} also carried out 2D MAD simulations with a prograde black hole spin of $0.9$, consistent with the modeling of the EHT. Through synchrotron GRRT calculations, they generated images at both 86~GHz and 230~GHz and from two different electron populations. They find that the diameter of the ring produced by non-thermally distributed (power-law) electrons is too small by at least $30\%$ compared to the observational value, while thermally distributed (Maxwell-Jüttner) electrons produce rings with diameters consistent with observations at both frequencies.
        
        In this work we modeled the event horizon and extended jet structure of M\,87 at submillimeter frequencies by combining GRMHD simulations and GRRT, exploring the impact of a mixed distribution of thermal and nonthermal electrons. In Sect. 2 we introduce our GRMHD setup. In Sect. 3 we characterize the GRRT calculations that we carried out to obtain our synthetic images. In Sect. 4 we show and analyze the output of our simulations. We not only study our data in the image plane but also focus on generating synthetic interferometric data to take the limited resolution of real observations into account. Finally, we discuss our results in Sect. 5. Throughout this work, we assumed a black hole mass ($m_{\rm BH}$) of $6.5\times10^9\,M_\odot$ and a luminosity distance ($d_L$) of $16.5\,\mathrm{Mpc}$. With these we computed the gravitational radius and time, which scale the dimensionless simulation output, $1\,r_g=Gm_{\rm BH}/d_Lc^2=3.8\,\mu$as and $1\,t_g=Gm_{\rm BH}/c^3=8.5\,h$.
        
        \section{GRMHD simulations}
        
        We simulated the accretion onto a rotating Kerr black hole with spin $a_\star=-0.5$ and its associated jet launching with the 3D GRMHD code \texttt{BHAC} \citep{porth2017,olivares2019}. Our choice of black hole spin was motivated by modeling studies of the linear polarization data of M\,87, wherein \citet{EHTh2021} found that MAD models with $a_\star=-0.5$ met their scoring-based criteria more than any other combination of parameters. Our simulation follows the same setup as \citet{fromm2022}, but we give a brief summary below.
        
        \texttt{BHAC} solves the equations of conservation of mass, conservation of energy-momentum, and the covariant Maxwell equations. We followed ideal, non-resistive magnetohydrodynamics, for which the stress-energy tensor takes the form
        
        \begin{equation}\label{eq:stressenergy}
                T^{\mu\nu}=\left(\rho h_\textrm{tot}\right)u^\mu u^\nu + \left(p+\frac{1}{2}b^2\right)g^{\mu \nu} - b^\mu b^\nu .
        \end{equation}
        
        \noindent Here, $\rho$ is the rest-mass density, $p$ is the gas pressure, $h_\textrm{tot}=h+b^2/\rho$ is the total specific enthalpy, $u^\mu$ is the four-velocity, $g_{\mu \nu}$ is the space-time four-metric and $b^\mu$ is the magnetic field four-vector. The specific enthalpy is computed by assuming an ideal gas equation of state,
        
        \begin{equation}\label{eq:enthalpy}
                h=1+\frac{\hat\gamma p}{\left(\hat\gamma-1\right)\rho},
        \end{equation}
        
        \noindent where $\hat\gamma$ is the adiabatic index. Throughout this work we used an adiabatic index of $\hat{\gamma}=4/3$. This value was chosen for the sake of consistency with recent \texttt{BHAC} simulations used by the EHT to model M\,87 \citep{EHTshadowII2025}. While this value results in more numerically robust simulations, two-temperature fluid simulations \citep{chael2025} as well as theoretical works based on thermodynamic principles \citep{gammie2025} find that a value closer to $5/3$ may be more appropriate to model single-fluid simulations. We defined the magnetization as $\sigma=b^2/\rho$ and the plasma beta as $\beta=2p/b^2$, calculated through the square of the magnetic field:
        
        \begin{equation}\label{eq:enthalpy}
                b^2=\frac{B^2}{\Gamma^2}+\left(B^i u_i\right)^2.
        \end{equation}
        
        \noindent Here, the Latin index $i$ refers only to the three spatial coordinates. $\Gamma$ is the Lorentz factor of the fluid, and $B^2=B^iB_i$, where $B^i$ is the three-vector of the magnetic field. Throughout this work we set $G=c=1$ and employed spherical Kerr-Schild coordinates, $(r,\theta,\phi)$. The outer radial edge of the simulation is located at $r=2500\,M$, and the inner one at $1.18r_\textrm{EH}\,M$, where $r_\textrm{EH}=1+\sqrt{1-a^2_\star}$ is the radius of the event horizon. Furthermore, we made use of three adaptive mesh refinement levels, which led to an effective resolution of $384\times192\times192$ cells in the $r$, $\theta,$ and $\phi$ directions.
        
        The initial setup of our simulations consists of a Fishbone-Moncrief torus \citep{FM1976} in hydrostatic equilibrium characterized by its inner edge, $r_\mathrm{in}=20\,M$, the location of the pressure maximum, $r_\mathrm{c}=40\,M$ and its specific angular momentum, $l=-u_\phi/u_t=6.88$. This torus is seeded with a weak poloidal magnetic field given by the vector potential,
        
        \begin{equation}\label{eq:vectorpotential}
                A_\phi = \max\left(\frac{\rho}{\rho_\mathrm{max}}\left(\frac{r}{r_\mathrm{in}}\right)^3\sin^3\theta\exp\left(\frac{-r}{400}\right)-0.2,0\right)
        ,\end{equation}
        
        \noindent and normalized to a plasma beta of $\beta=100$ at $r_\textrm{in}$.
        
        The initial hydrostatic equilibrium is perturbed by a small pressure variation that triggers the formation of the magnetorotational instability, kick-starting the accretion. During the course of the simulation up to $t=30000\,M,$ we monitored the mass accretion rate ($\dot{M}$) and the accreted magnetic flux, $\phi_{\rm BH}$. These were evaluated at the event horizon and defined as
        
        \begin{equation}
                \dot{M}=\int^{2\pi}_0\int^{\pi}_0\rho u^r\sqrt{-g}d\theta d\phi,
                \label{eq:mdot}
        \end{equation}
        
        \begin{equation}
                \phi_{\rm{BH}}=\frac{1}{2}\int^{2\pi}_0\int^{\pi}_0|B^r|\sqrt{-g}d\theta d\phi.
                \label{eq:phi}
        \end{equation}
        \begin{figure}[h!]
                \includegraphics[width=9cm]{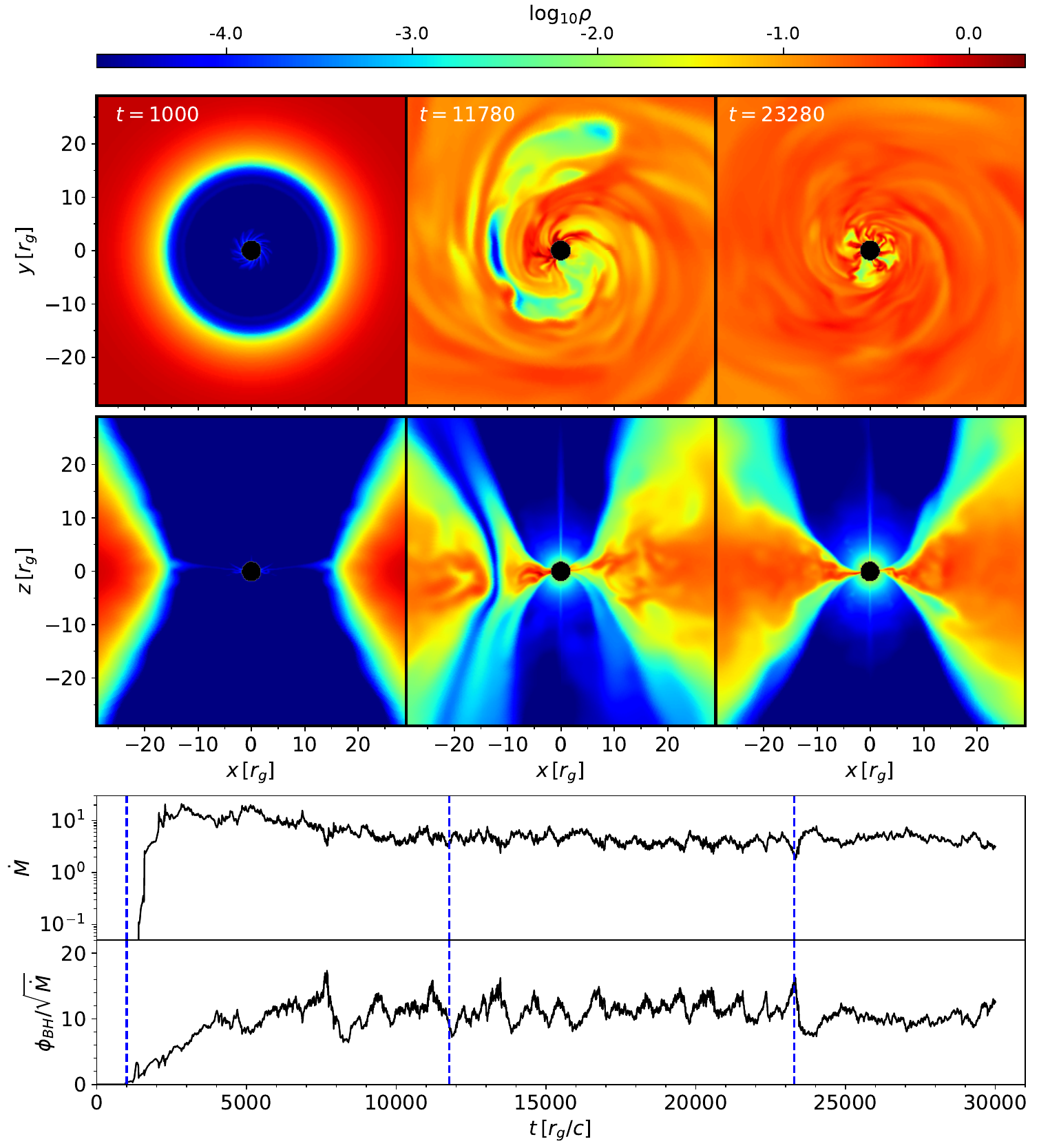}
                \caption{Evolution of our GRMHD simulation over time. The top and middle panels show the logarithm of the density in the equatorial and a meridional plane, respectively, for three different times. The bottom panel shows the temporal evolution of the mass accretion rate ($\dot{M}$) and the MAD parameter, $\Phi=\phi_{\rm BH}/\sqrt{\dot{M}}$. The vertical dashed blue lines mark the time frames of the density plots. }
                \label{fig:grmhd}
        \end{figure}
        
        Figure~\ref{fig:grmhd} showcases our initial setup and its evolution over time. The six panels on top display density maps of the centralmost region of our simulation. The top and bottom row of images display the equatorial and a meridional plane, respectively, at three different times during the course of the simulation. The leftmost panels show the density shortly after the start of the simulation, when the accretion process is not yet fully established, evidenced by the low-density gap between the central black hole and the high density innermost region of the torus. The two advanced time steps show both the accretion onto the black hole and the launching of a jet and a counter-jet, which are seen as low density funnels in the meridional plane. The central column shows a low density arc of matter being ejected from the direct vicinity of the black hole, a common event in MADs and one of their characteristic features. During these MAD events, the accumulated magnetic pressure pushes part of the disk outward, leading to low-density and highly magnetized regions known as MAD bubbles, which are then sheared out in the accretion flow and disappear. The bottom panels show the evolution of the mass accretion rate ($\dot{M}$) and the MAD parameter, $\Phi=\phi_\textrm{BH}/\sqrt{\dot{M}}$. The latter oscillates around a value of 10, indicating that a MAD state has been reached as defined by \citet{tchekhovskoy2011}.
        
        \section{Radiative transfer}
        
        We carried out the radiative transfer calculations with the GRRT code \texttt{BHOSS} \citep{younsi2012,younsi2016,younsi2020}. The output of \texttt{BHAC} is straightforwardly used as input for the post-processing code \texttt{BHOSS}, which solves the equations of covariant GRRT,
        
        \begin{equation}
                \frac{\textrm{d}\tau_\nu}{\textrm{d}\lambda}=\gamma^{-1}\alpha_{0,\nu},
                \label{eq:opticaldepth}
        \end{equation}
        
        \begin{equation}
                \frac{\textrm{d}\mathcal{I}}{\textrm{d}\lambda}=\gamma^{-1}\left(\frac{j_{0,\nu}}{\nu^3}\right)\mathrm{exp}(-\tau_\nu),
                \label{eq:intensity}
        \end{equation}
        
        \noindent alongside the affine parameter $\lambda$ in environments with a strong gravity. Here the optical depth is defined as $\tau_\nu= \int \alpha_\nu \,\textrm{d}\lambda$, where  $\alpha_\nu$ is the absorption coefficient (or absorptivity) of the plasma that comprises the medium. $j_\nu$ is the emission coefficient (or emissivity) of the plasma and the subscript $0$ indicates that it is the value in its local rest frame. The energy shift with respect to the observer's frame is defined as $\gamma^{-1}=\nu_0/\nu$, and the Lorentz-invariant intensity $\mathcal{I}$ is related to the specific intensity $I_\nu$ as $\mathcal{I}=I_\nu/\nu^3$.
        
        We largely followed the same method as \citet{fromm2022}, where the GRRT calculations are explained in more detail, but we give here a summary for the sake of completeness. As the power associated with synchrotron radiation depends on the particle mass as $P\propto m^{-2}$ \citep{rybicki1986}, and $m_\textrm{e}\ll m_\textrm{p}$, the majority of the synchrotron emission comes from the electrons in the plasma. To calculate the absorption and emission coefficients of Eqs.~\ref{eq:opticaldepth} and \ref{eq:intensity}, we needed to establish the properties of the electrons, whose distribution function will depend on the dimensionless electron temperature $\Theta_\textrm{e}=kT_\textrm{e}/m_\textrm{e}c^2 $.
        
        The GRMHD simulations carried out in Sect.~2 are single-fluid simulations, and thus the resulting temperature is only that of the protons in the plasma. We calculated the electron temperature $\Theta_\textrm{e}$ from the proton temperature following the R-$\beta$ model of \citet{moscibrodzka2016},
        
        \begin{equation}
                \Theta_\textrm{e}=\frac{pm_\textrm{p}/m_\textrm{e}}{\rho T_\textrm{ratio}}.
        \end{equation}
        
        The ratio between proton and electron temperature is regulated by the plasma beta $\beta$, following the formula
        
        \begin{equation}
                T_\textrm{ratio}\equiv\frac{T_\textrm{p}}{T_\textrm{e}}=\frac{R_\textrm{low}+R_\textrm{high}\beta^2}{1+\beta^2}.
        \end{equation}
        
        \noindent This model introduces the parameters $R_\textrm{low}$ and $R_\textrm{high}$, which regulate the temperature ratio in the regions where the gas pressure (disk) and the magnetic pressure (jet) dominate, respectively. 
        
        In this work we used a hybrid electron distribution function (eDF), consisting of a mixed thermal and nonthermal distribution. In this type of astrophysical scenario, electrons are expected to be accelerated to nonthermal distributions through mechanisms such as turbulence, magnetic reconnection, or shocks. Here, we parametrized the nonthermal electrons via a kappa distribution \citep{davelaar2019,cruzosorio2022,fromm2022,Zhang2024}. The kappa eDF takes a thermal-like shape for particles with small electron Lorentz factors, $\Gamma_\textrm{e}$, but transitions to a nonthermal power-law tail for large $\Gamma_\textrm{e}$ \citep[for details, see][]{pandya2016}. The kappa eDF is parametrized by the slope of its nonthermal tail, i.e., the kappa parameter, $\kappa$, and its width, $w$. Following \citet{davelaar2019} we calculated $w$ as        
        \begin{equation}
                w=\frac{\kappa-3}{\kappa}\Theta_e + \frac{\epsilon}{2}\left[1+\tanh{\left(r-r_{\textrm{inj}}\right)}\right]\frac{\kappa-3}{6\kappa}\frac{m_\textrm{p}}{m_\textrm{e}}\sigma. 
                \label{eq:kappaw}
        \end{equation}

        \begin{figure*}[t]
                \includegraphics[width=0.99\textwidth]{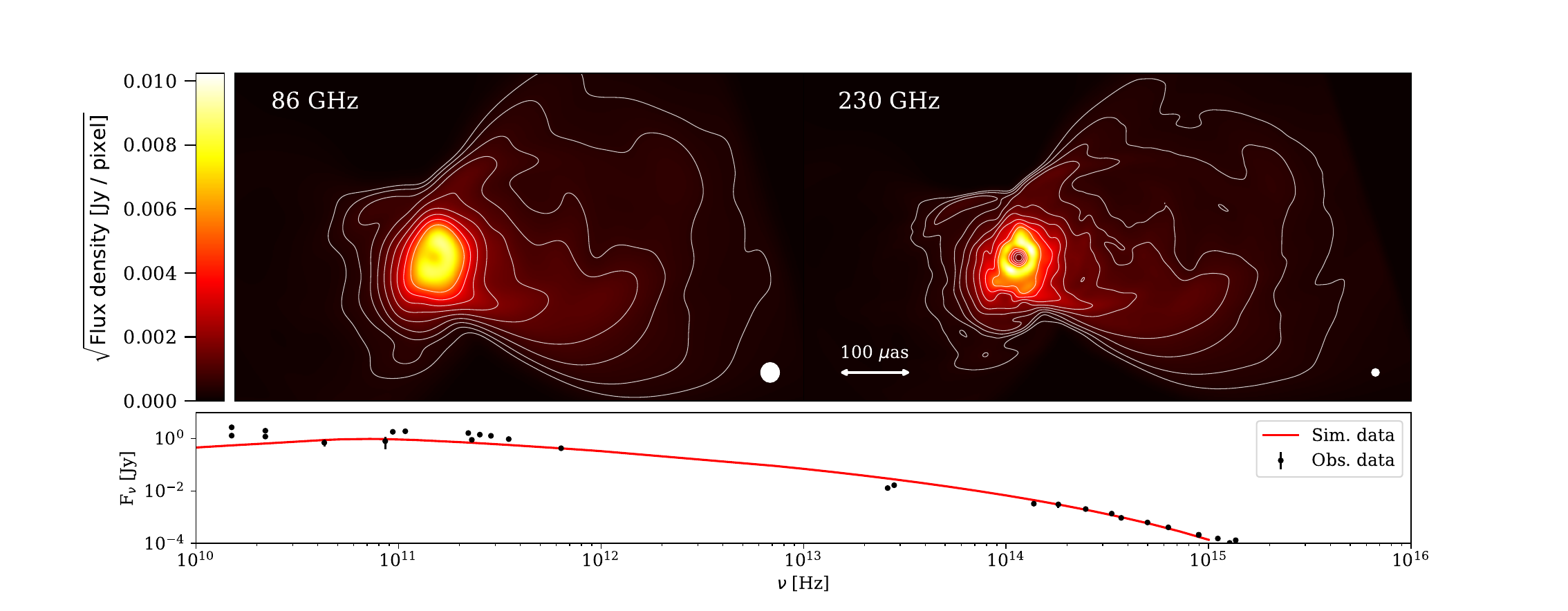}
                \centering
                \caption{Results of the GRRT calculations at $t=26000\,M$. The images show the flux density at 86\,GHz and 230\,GHz. Notice that the images are convolved with half of the nominal resolution of the GMVA (27$\,\mu$as) and the EHT (10$\,\mu$as), indicated by the white circle at the bottom right. The contours begin at $1/1000$ of the maximum flux density and increase by factors of two. In the bottom panel we display the broadband radio spectrum of our model. The observational flux values are taken from \citet{prieto2016}, \citet{perlman2001}, \citet{whysong2004}, \citet{hada2017}, and \citet{lister2018}.}
                \label{fig:mosaic}
        \end{figure*}
        
        \noindent We assumed that the energy in the kappa distribution is coupled to a thermal value, and we included some fraction ($\epsilon$) of the magnetic energy. This additional energy mimics the acceleration of particles, for example via magnetic reconnection after some distance ($r_\mathrm{inj}$).
        
        The value of the $\kappa$ parameter needs to be evaluated everywhere in the plasma. Past works, such as \citet{davelaar2018}, found that a fixed value of $\kappa$ could not simultaneously fit the known spectral energy distribution and spectral index values of Sgr\,A$^\star$. They initially fixed this by reducing the number of accelerated electrons through an additional free parameter, a nonthermal/thermal electron fraction. \citet{davelaar2019} improved on the model for M\,87, parameterizing electron acceleration from first principles through the results from the particle-in-cell relativistic reconnection simulations of \citet[see also Sect. 4.2. of \citealt{ehtsgra2022}]{ball2018}. We used the same model of electron acceleration in our work. \citet{ball2018} fit the electron power-law slope to the plasma beta and magnetization and obtained the analytical expression
        
        \begin{equation}
                \kappa=2.8+0.7\sigma^{-1/2}+3.7\sigma^{-0.19}\tanh{\left(23.4\sigma^{0.26}\beta\right)}.
                \label{eq:kappapara}
        \end{equation}
        
        Based on the calculations of \citet{pandya2016} for the relative fitting errors of the kappa distribution, we restricted our interval to $3<\kappa\leq8$. Up until this point, the GRRT calculations followed those of \citet{fromm2022}, which computed the emission and absorption coefficients following a kappa distribution within this range and a thermal distribution outside of it. We introduced a change in this step, still following the equations of \citet{pandya2016} for the $j_\nu$ and $\alpha_\nu$ coefficients. While we continued to use the coefficients of a thermal distribution, $j_{\nu,\textrm{thermal}}$ and $\alpha_{\nu,\textrm{thermal}}$, to carry out the GRRT calculations outside of our $\kappa$ range, we introduced here a mixed efficiency model within the range        
        \begin{equation}
                j_{\nu,\textrm{tot}}=(1-\tilde{\epsilon})j_{\nu,\textrm{thermal}}+\tilde{\epsilon}j_{\nu,\kappa}.
                \label{eq:emissivity}
        \end{equation}
        
        \noindent The absorption coefficients are similarly computed, and the thermal and nonthermal contributions are distributed through a new magnetic efficiency parameter, $\epsilon_\textrm{eff}$, following the equation

        \begin{equation}
                \tilde{\epsilon}=\epsilon_\textrm{eff}\left[1-e^{-1/\beta^2}\right]\left[1-e^{-\left(\sigma/\sigma_\textrm{min}\right)^2}\right].
                \label{eq:epsilon}
        \end{equation}
        
        \setlength{\tabcolsep}{5pt}
        \renewcommand{\arraystretch}{1}
        \begin{table}[h]
                \centering
                \begin{threeparttable}
                        \caption{Parameters of the GRRT calculations in this paper.}
                        \label{table:grrtparams}
                        \begin{tabular}{ccccccccc}
                                \hline \hline \\[-0.9em]
                                $R_\textrm{low}$    & $R_\textrm{high}$  & $\sigma_{\rm min}$ & $\sigma_\textrm{cut}$  & $\epsilon$ & $\epsilon_\textrm{eff}$ & $r_\textrm{inj}$ & $\dot M$ [$m_\odot$/yr] & $\vartheta$ [$^\circ$] \\ \\[-0.9em] \hline \\[-0.8em]
                                3 & 160  & $10^{-5}$ & 5 & 0.4 & 0.8 & 10 & 1.25$\times10^{-4}$ & 17 \\ \\[-0.9em] \hline 
                        \end{tabular}
                        
                        \begin{tablenotes}
                                \small
                                \item \textbf{Notes:} Left to right: $R$ parameters for the electron temperature, lower and upper magnetization cutoffs, $\epsilon$ parameters for the electron acceleration efficiency, injection radius of nonthermal particles, total mass accretion rate, viewing angle.
                        \end{tablenotes}
                \end{threeparttable}
        \end{table}
        
        As the initial ambient medium of our simulations is unmagnetized, and a large fraction of it will remain unmagnetized over the course of the simulation, Eq.~\ref{eq:epsilon} would return not-a-number values in the GRRT calculations. To avoid these numerical issues, we introduced  $\sigma_\textrm{min}$ as a small cutoff value, well below the initial magnetization of the torus. Appendix~\ref{sec:emissappendix} contains a visual of the value that $\tilde{\epsilon}$ takes on the disk and jet regions of our simulation.
        
        Our final model leaves us with a large parameter space to study. While \citet{fromm2022} carried out a careful study of the impact of these parameters on the image properties and spectra, in this study we fixed the parameters to the values given by Table~\ref{table:grrtparams}. These were chosen for their ability to retrieve an extended jet structure and to fit the high-frequency part of the spectrum. Specifically, $\sigma_\textrm{cut}$ is a cutoff value for the maximum magnetization that will be included in the GRRT. High magnetization values can be found in the funnel of the jet, but this region is known to be affected by the floor-model used to increase the stability of the code. We set $\sigma_\textrm{cut}=5$ to exclude this region during the GRRT calculations. This value is larger than the standard value of 1 \citep{EHTshadowII2025}, meaning our model includes more emission from the jet funnel than previous studies. \citet{fromm2022} found that $\sigma_\textrm{cut}\geq3$ resulted in an increased jet width and in the presence of a central ridge, likely due to the previously mentioned floor-model, as more magnetized, hot plasma is being included in the jet sheath with increasing $\sigma_\textrm{cut}$. The mass accretion rate was calculated via a minimization process to match the total flux at 86~GHz to observational data, 0.8~Jy, as observed by \citet{lu2023}.

        \section{Results}
        
        The GRRT simulations were carried out from $t=20000\,M$ to $t=29990\,M$, with a cadence of $t_{\rm cad}=10\,M\approx3.7\,\mathrm{days}$. We computed the emission over a frequency range of $1\times10^9$\,Hz to $1\times10^{15}$\,Hz, spaced logarithmically into 50 bins. We then generated images from the 86~GHz and 230~GHz data in a $2048\times2048$ pixel grid covering 200~M. This corresponds to an angular size of around 760$\,\mu$as using the black hole mass and distance of M\,87.
        We used the fast-light approximation, ignoring the finite speed of light during the radiative transfer calculations.

        The top panels of Fig.~\ref{fig:mosaic} show the output of our GRRT calculations at $t=26000\,M$ for 86\,GHz and 230\,GHz. The images are blurred with half of the nominal resolution of the GMVA and the EHT, indicated by the white circle at the bottom right. They are dominated by a bright central ring-like structure and display an edge-brightened jet. The bottom panel shows the broadband radio spectrum of M\,87 alongside the spectrum obtained from our model (solid red line).
        
        \begin{figure}[h!]
                \includegraphics[width=9cm]{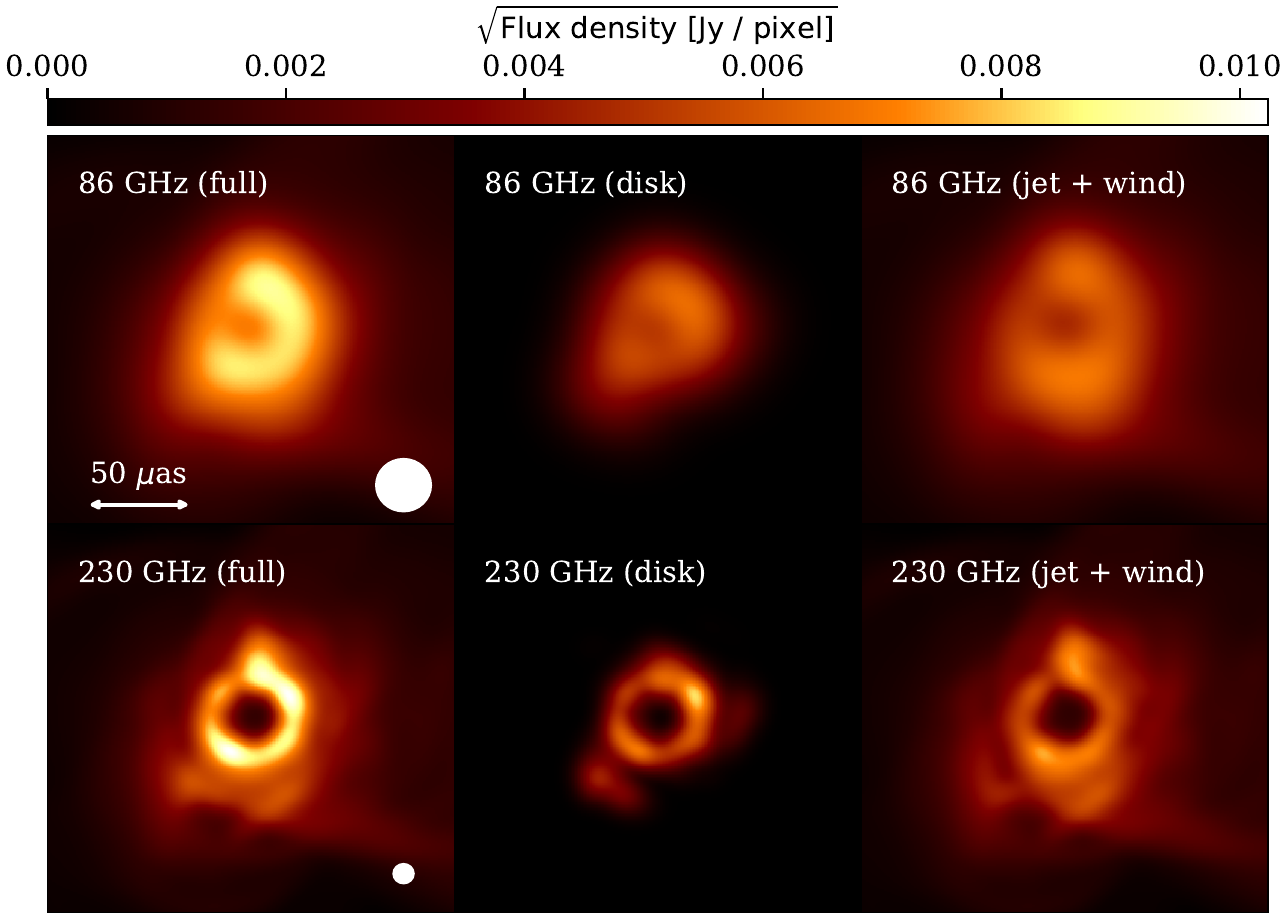}
                \caption{Structural decomposition of the horizon scale structure in our model at $t=26000\,M$. The panels show from left to right the total, disk, and jet plus wind contributions for 86\,GHz (top row) and 230\,GHz (bottom row).}
                \label{fig:split}
        \end{figure}
        
        In Fig.~\ref{fig:split} we zoom into the horizon region of Fig.~\ref{fig:mosaic}. We additionally decompose the image into the emission originating from the disk and from the jet, to better understand their contribution to the total image. For this purpose, during the GRRT calculations we set the emission coefficients to zero in either the jet or disk while keeping the absorption coefficients unchanged. We used the Bernoulli parameter $-hu_t=1.02$ to separate the disk and jet, marking the transition between the bound gas of the disk and the unbound gas of the jet \citep{porth2019,Moriyama2024}. The decomposition shows that the innermost region is dominated by emission from the disk, but there is some contribution from the jet foot-points at larger radii. It is interesting to remark that the bright spots seen in the disk and jet decompositions do not necessarily coincide, which indicates that on horizon scales we see the variability in both the accretion flow and in the jet formation zone. We provide further details and discussion on temporal variation in Sect. \ref{sec:disc} and Appendix~\ref{sec:fittingresults}. 
        
        \begin{figure}[h!]
                
                \includegraphics[width=0.5\textwidth]{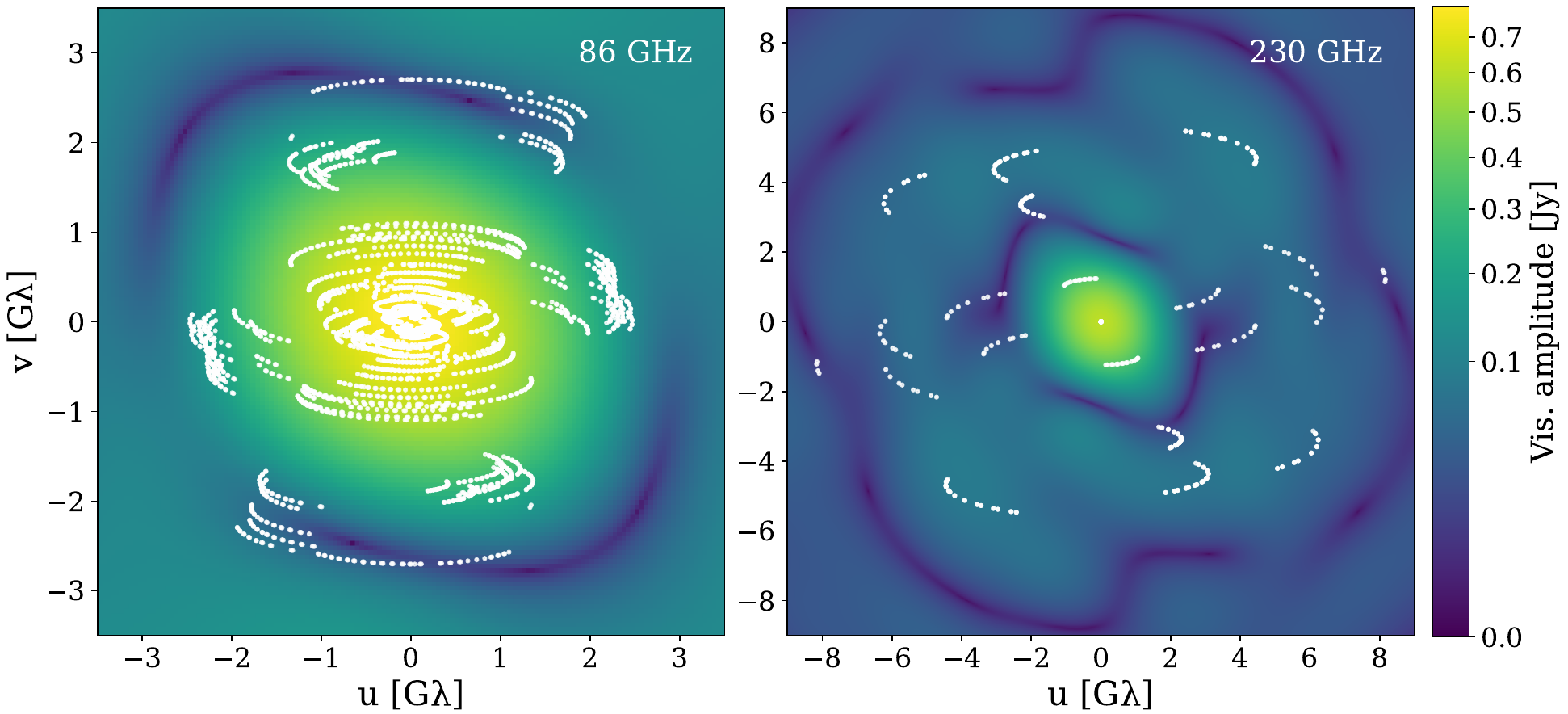}
                \centering
                \caption{Fourier transforms of the central regions of our simulated data at $t=24500\,M$. Plotted on top is the $u-v$ coverage at 86\,GHz and 230\,GHz used for the generation of synthetic observations, using the array configurations presented in \citet{lu2023} and \citet{EHTa2019}, respectively. Note the different scales in the two plots.}
                \label{fig:uvcoverage}
        \end{figure}

        \begin{table}[h]
                \centering
                \caption{Antennas involved in the observations used to generate synthetic data and their nominal sensitivities.}
                \label{table:sefd}
                \begin{threeparttable}
                        \begin{tabular}{cccc}
                                \hline \hline \\[-0.9em]
                                \multicolumn{2}{c}{86 GHz Antennas} & \multicolumn{2}{c}{230 GHz Antennas} \\ \hline \\[-0.9em]
                                Antenna & SEFD [Jy] & Antenna & SEFD [Jy] \\ \\[-0.9em] \hline \\[-0.8em]
                                VLBA\tnote{a} & 4100 &  APEX & 4700 \\
                                Effelsberg & 1000 & JCMT & 10500 \\
                                Green Bank & 137 & LMT & 4500 \\
                                Metsähovi & 15000 & SMT & 17100 \\
                                Yebes & 990 & SMA & 6200 \\
                                Onsala & 2650 & Pico Veleta & 1900 \\
                                ALMA & 68 & ALMA & 74 \\
                                Pico Veleta & 654 & - & - \\
                                GLT & 5312 & - & - \\ \\[-0.9em] \hline 
                        \end{tabular}
                        \begin{tablenotes}
                                \small
                                \item[a] Consists of the 8 antennas that function at 86~GHz.
                        \end{tablenotes}
                \end{threeparttable}
        \end{table}

        In order to compare our numerical results to observations, we computed synthetic visibilities from our images and perform the analysis in the Fourier plane. To do this, we used the \texttt{ehtim} package \citep{chael2018} to Fourier transform and sample all of our images with the baselines of real observations. This was done through the equation
        
        \begin{equation}\label{eq:visibilities}
                V_{ij} =\int\int I(x,y)e^{-2\pi \textrm{i}(ux+vy)}\textrm{d}x\textrm{d}y,
        \end{equation}
        
        \noindent where a baseline formed by the $(i,j)$ stations detects a complex visibility $V_{ij}$. Figure~\ref{fig:uvcoverage} shows the $u-v$ coverage used in this paper, comprising every $(u,v)$ pair. We sampled every 86~GHz image with the baselines of \citet{lu2023} and added the errors they calculated to the visibilities. We similarly produced synthetic visibilities for our one thousand 230~GHz images with the baselines of the 2017 observation of M\,87 by the EHT \footnote{\url{https://eventhorizontelescope.org/for-astronomers/data}}. The stations that participated in both observations, and their respective sensitivities in system-equivalent flux density (SEFD), are given in Table~\ref{table:sefd}. Throughout this paper we refer to the synthetic visibilities created from our GRMHD and GRRT models as ``data.'' However, our ``data'' should not be confused with real observational data of the GMVA and the EHT.
        
        We conducted our study of the ring morphology over two quantities: the visibility amplitudes and the closure phases. Closure phases were defined as the phase $\psi_\mathrm{C}$ of the product of three visibilities in a triangle formed by three stations,
        
        \begin{equation}
                \psi_{\mathrm{C},ijk}=\mathrm{arg}(V_{ij}V_{jk}V_{ki}).
        \end{equation}
        
        \noindent Being the product of a triangle of baselines makes closure phases robust to station-based calibration errors, making them useful in radio interferometry  \citep{chael2018}. Furthermore, nonzero closure phases are indicative of an azimuthally asymmetric structure, and have been observed for M\,87 at 230~GHz \citep{EHTa2019}. Indeed, modeling work done by the EHT supports the necessity of fitting azimuthally asymmetric models to the data \citep{EHTf2019}. 
        
        It is relatively straightforward to fit symmetric models such as Gaussians, thin rings and disks to the data. The Fourier transforms of these geometries are proportional, respectively, to a Gaussian, a Bessel function of order zero and of order one. Furthermore, due to being real-valued everywhere, the closure phases can only take on the values of $0^{\circ}$ or $180^{\circ}$. A thick ring, modeled by \texttt{ehtim} as a thin ring blurred by a Gaussian kernel, is similarly symmetric but may be able to successfully retrieve the underlying structures. \citet{lu2023} fit all of these symmetric models to the 86~GHz, 14-15 April 2018 observation of M\,87, and find a thick ring to be the best-suited model to fit the data and retrieve the ring diameter. While we used an asymmetric model to fit our data at both frequencies, as explained in the next section, we explore the performance of these symmetric models on our time-variable data in Appendix~\ref{sec:symmmodels}.

        \subsection{86~GHz fitting}
        
        The EHT used Bayesian evidence to evaluate in depth how different geometric models of increasing complexity fit 230~GHz M\,87 data from 2017 and 2018 \citep{EHTf2019,EHTshadowI2024}. They arrived at the conclusions that (i)~ring-like models overall fit the data much better than non-ring models, with the best-performing non-ring model consisting of three elliptical Gaussians, and (ii)~the symmetric models are the worst performers by several orders of magnitude. Their best performing models contain upward of 20 free parameters. Fitting models with this many degrees of freedom is challenging since they require robust bounds, which can be difficult to mantain over a large amount of time steps with a degree of variability that is unknown in advance, as is our case. We decided to use the model that contains the fewest degrees of freedom while still performing well: a thick $m$-ring with $m=2$. This model is described in polar coordinates $(r,\phi)$ by the equation \citep[see][]{johson2020}
        
        \begin{equation}\label{eq:mringimage}
                I_\mathrm{ring}(r,\phi)=\frac{F_0}{\pi d}\delta(r-d/2)\frac{4\ln2}{\pi\alpha^2}e^{-\frac{4\ln2r^2}{\alpha^2}}\sum^{m}_{k=-m}\beta_ke^{ik\phi}.
        \end{equation}
        
        \noindent Here, the first term containing the delta function is the expression for an infinitesimally thin ring of total flux $F_0$ and diameter $d$. The exponential term corresponds to a Gaussian that blurs the ring and turns it into a thick ring of thickness $\alpha$. Finally, the ring is modulated by a sum of Fourier modes with complex coefficients $\beta_k$, not to be confused with the plasma beta. It is this term that confers an azimuthal structure to the ring, specifically up to a quadrupole in the case of our $m=2$ model. The visibilities of the thick $m$-ring model are given by the Fourier transform of Eq.~\ref{eq:mringimage},
        
        \begin{equation}\label{eq:mringvis}
                V_\mathrm{ring}(u,\phi_u)=F_0e^{-\frac{\pi^2\alpha^2u^2}{4\ln2}}\sum^{m}_{k=-m}\beta_kJ_k(\pi ud)e^{ik(\phi_u-\pi/2)}.
        \end{equation}
        
\begin{figure}[h!]
                \includegraphics[width=9cm]{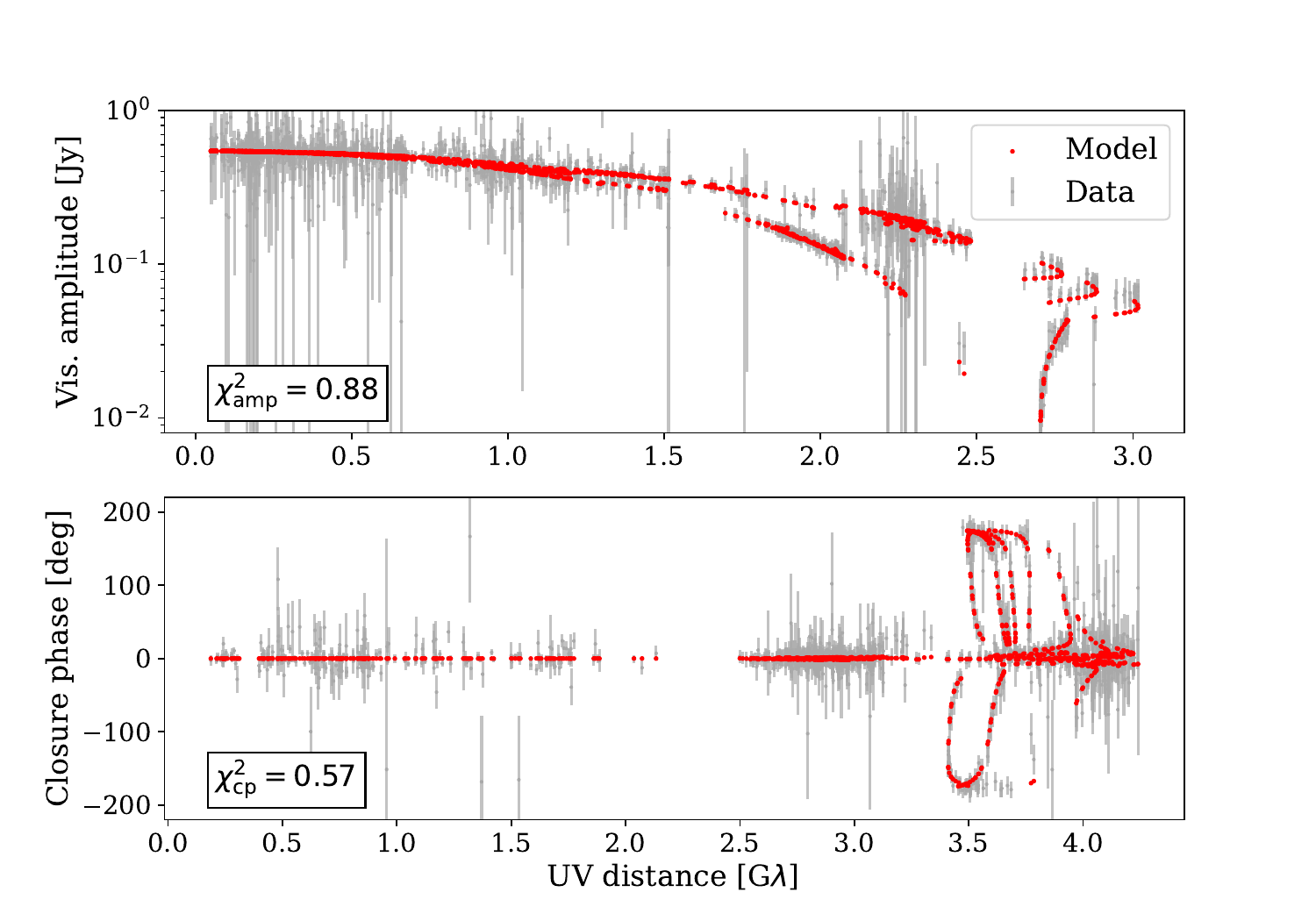}
                \caption{Visibility amplitudes and closure phases over uv distance for the time step $t=28600\,M$ at 86~GHz. Overlaid in red is the result of fitting a thick $m$-ring model to the data. The $\chi^2$ values have been calculated taking an additional 10\% error  into account  for ALMA visibilities.}
                \label{fig:mringvis}
        \end{figure}

       \noindent $J_k$ is the Bessel function of order $k$, while $(u,\phi_u)$ are the polar coordinates in the interferometric plane. The sum of $m$-modes distinguishes this model from symmetric ones, as this function is complex-valued and thus can have closure phases that are not $0^\circ$ or $180^\circ$ everywhere. This can be seen in the bottom panel of Fig.~\ref{fig:mringvis}, which shows the closure phases of a model (red) fit to the data of a specific time step (gray). The nonzero closure phases in the long baselines are indicative of azimuthal structure, and they are successfully modeled by Eq.~\ref{eq:mringvis}. The top panel, which shows the visibility amplitudes, also highlights the necessity of the last term in Eq.~\ref{eq:mringvis} to fit many of our time steps. The sum of Bessel functions can retrieve the structures observed in the middle and long baselines, where the same uv distance can contain several different values of the visibility amplitudes. This implies an asymmetry in the visibilities observed by baselines of similar length, pointing toward, for example, a north-south and east-west asymmetry. 
        
        We carried out the fitting shown in Fig.~\ref{fig:mringvis} for all of our time steps at 86~GHz. We used the baseline and error information of \citet{lu2023} to sample the Fourier transform of each of our images and, following their work, average the visibilities every 420 seconds. This operation was done on the central 100$\,\mu$as of each image so as to not include the bulk of the extended jet emission in the dataset, as explained in Appendix~\ref{sec:fieldofview}. To fit the resulting visibilities, we used the modeling capabilities of the Python package \texttt{ehtim} \citep{chael2016,chael2018}. We chose to use the minimizer function ``dynamic nested sampling'' of the \texttt{dynesty} package \citep{speagle2020,skilling2004,skilling2006,higson2019}. This evaluates the Bayesian evidence of a model and allows us to examine the final posteriors of the fit parameters. We show one set of such posteriors in Appendix~\ref{sec:fittingresults}.
        
        We evaluated the goodness of fit of our models to the data through the time-averaged mean squared residuals, $\chi^2$, for the amplitude and closure phases. The $\chi^2$ are computed as        
        \begin{equation}\label{eq:chisqamp}
                \chi^2_\mathrm{amp} = \frac{1}{N_\mathrm{V}}\sum{\frac{(A_{obs}-A_{mod})^2}{\sigma_A^2}},
        \end{equation}
        
        \begin{equation}\label{eq:chisqcp}
                \chi^2_\mathrm{cp} = \frac{2}{N_{\psi_\mathrm{C}}}\sum{\frac{1-\cos{({\psi_\mathrm{C,obs}}-{\psi_\mathrm{C,mod}})}}{\sigma_{\psi_\mathrm{C}}^2}}.
        \end{equation}
        
        \noindent To understand whether the thick $m$-ring model is adequate to fit our large dataset, we first ran a series of tests on a smaller subset of our data, which can be found in Appendix~\ref{sec:symmmodels}. A weak fit will not retrieve reliable parameters from the data, as features such as the diameter or thickness of a ring are related to the nulls in the visibilities. We find that a thick $m$-ring model provides a reasonably good fit for most time steps, despite its relatively low number of degrees of freedom.  We then used this $m=2$ $m$-ring model for our larger 86~GHz dataset and added an error of 10\% of the visibility amplitude to ALMA visibilities. The GMVA together with ALMA and the GLT comprise a heterogeneous array, with ALMA data having a much higher signal-to-noise ratio than other stations. Thus, adding a fractional error to the visibilities of those baselines aids in the convergence of the fitting routine, which would be otherwise dominated by the small error bars of ALMA data. This adjustment of ALMA uncertainties has been used on the 230~GHz EHT observations~\citep{EHTd2019,EHTshadowI2024}. \citet{EHTd2019} tests adjusting the weights of ALMA visibilities to values between 0.01 and 1.0 with a similar motivation of avoiding data from these baselines dominating the phase and amplitude correction. For the 2017 230~GHz data, this would correspond to an added fractional error of $0\%-40\%$ of the visibility amplitudes. We find that for our 86~GHz data, a 10$\%$ error returns satisfactory results in aiding convergence without having ALMA data dominate the geometric fitting.

        \subsection{230~GHz fitting}
        
        The 230~GHz visibilities were computed from the baselines and errors of \citet{EHTa2019}. The real April 2017 EHT observations had a variable error budget; the Large Millimeter Telescope (LMT) had a particularly high one due to poor amplitude calibration \citep[see][]{eht2019c}. We assigned a 10\% fractional error to all stations so as to not excessively down-weight any particular station.
        
        We first carried out a study of the fitting models on a smaller subset of data, in the same time range as the one chosen for 86~GHz. We found that the $m=2$ $m$-ring could not satisfactorily fit the visibilities at 230~GHz, and decide on a more complex geometric model. Following the example of \citet{EHTf2019,EHTa2022}, we added two asymmetric Gaussians, with free major and minor full widths at half maximum, position angle, flux, and $(x,y)$ coordinates. This raises the number of parameters from 7 to 19. These free Gaussians can model generic features of the emission surrounding the central ring, which becomes distinct due to the decreasing optical thickness. More information can be found in Appendix~\ref{sec:230ghzmodel}.  We increased the field of view (FOV) to 140$\,\mu$as to compute the visibilities. We find this change does not significantly affect the diameter or thickness retrieved from the model fitting, unlike the 86~GHz case, as shown in Appendix~\ref{sec:fieldofview}. However, the increased FOV reduced the clipping of the Gaussian features out of frame, helping better locate their centroids.
        
        \section{Discussion}
        \label{sec:disc}
        \subsection{Ring size}
        
        \begin{figure}[h!]
                \includegraphics[width=9cm]{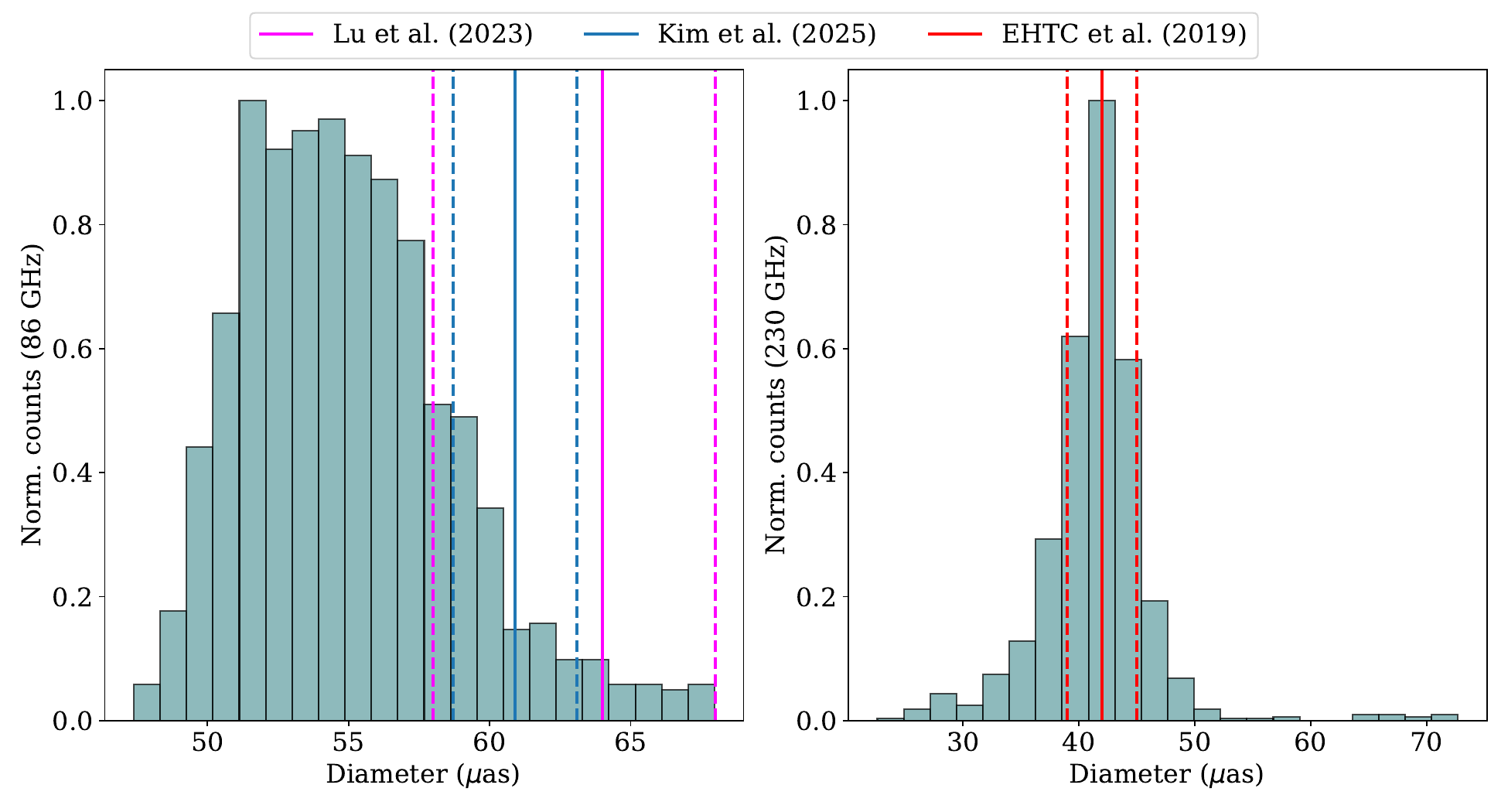}
                \caption{Normalized distribution of the diameters resulting from fitting our models to the 86~GHz and 230~GHz datasets. The other parameters can be found in Appendix~\ref{sec:fittingresults}. The vertical lines show the central observational values (solid) as well as upper and lower range (dashed) of the ring diameter at both frequencies. We use the range found by \citet{EHTa2019} at 230~GHz. At 86~GHz, we plot the range found by \citet{lu2023} and the range found by \citet{Kim2025} using the \texttt{resolve} algorithm.}
                \label{fig:hist}
        \end{figure}
        
        We show the distribution of diameters resulting from applying the fitting procedure to all one thousand of our time steps in Fig.~\ref{fig:hist}. We provide the information pertaining to the total flux and ring thickness in Appendix~\ref{sec:fittingresults}.
        
        We calculated mean values of $55.0\pm3.7\,\mu$as for the 86~GHz ring and $41.4\pm4.9\,\mu$as for the 230~GHz ring. Using the median values, and the distance to the Q1 and Q3 quartiles as lower and upper errors, we calculated $54.6^{+2.7}_{-2.4}\,\mu$as at 86~GHz and $41.6^{+1.8}_{-2.3}\,\mu$as at 230~GHz. While the 230~GHz distribution is strongly peaked, the higher number of degrees of freedom in the model result in more outliers, which affects the standard deviation. We use the median and interquartile values in the rest of the paper. 
        
        Up until here, we carried out our fitting procedure on GRRT data where the black hole mass and distance to us had the central values of $m_\mathrm{BH}=6.5\times 10^9M_\odot$ and $d_L=16.5\,\mathrm{Mpc}$, respectively. This matches the central value of angular gravitational radius estimated by \citet{EHTa2019}, $GM/Dc^2=3.8\pm0.4\,\mu$as.

        \begin{figure}[h!]
                \includegraphics[width=9cm]{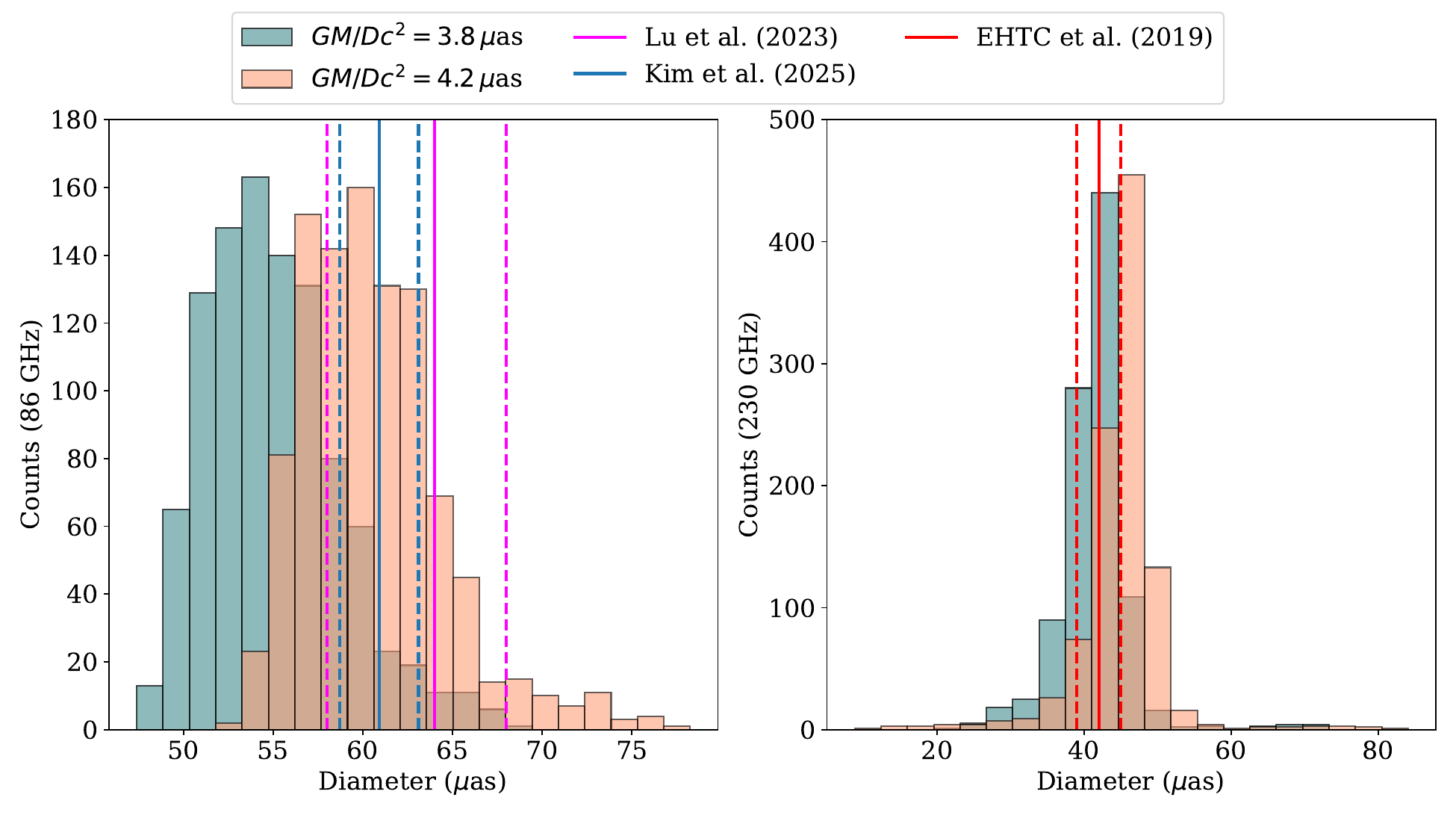}
                \caption{Distribution of the diameters resulting from fitting our models to the 86~GHz and 230~GHz datasets. The turquoise and orange histograms show, respectively, the result of model fitting to GRRT data with an angular scale of 3.8$\,\mu$as and 4.2\,$\mu$as. The vertical lines show the observational range for the diameter of the ring, using the same values as Fig.~\ref{fig:hist}.}
                \label{fig:highMD}
        \end{figure}
        
        We repeated our fitting procedure taking into account this $0.4\,\mu$as uncertainty in our current values for the mass of the M\,87 black hole and its distance to us. We incorporated this $\approx10\%$ error in the angular size of our image, increasing it to match the $GM/Dc^2=4.2\,\mu$as value and repeating the analysis to estimate an upper bound for our parameters. Increasing the FOV to match the angular size, we find the new fittings to be largely consistent with the old ones in the image plane, as expected. The $GM/Dc^2=4.2\,\mu$as data yield median diameters of $60.0^{+2.8}_{-2.4}\,\mu$as at 86~GHz and $45.4^{+1.9}_{-1.8}\,\mu$as at 230\,GHz. While the latter is on the upper error range of the observational value observed by the EHT, $42\pm3\,\mu$as, this shows the impact on the ring size of the uncertainties in the angular scale, that is, in the mass and distance of the black hole.
        
        While the GRRT images computed at 86~GHz with an angular scale of 4.2$\,\mu$as produce a median ring size closer to the central observational values of $64^{+4}_{-8}\,\mu$as and $60.9\pm2.2\,\mu$as, the upper bound of $54.6^{+2.7}_{-2.4}\,\mu$as from the 3.8$\,\mu$as images overlaps with the lower bounds of the observational value of \citet{lu2023}. Even at this angular scale, a significant number of time steps appear to fall within the error range of the observed values, as seen in Fig.~\ref{fig:hist}. We find diameters that fall within the observational range of \citet{lu2023} for around one third of the time steps. We explore the nature of these in Sect.~\ref{sec:nonthermal}.
        
        \subsection{Azimuthal structure}
        \label{sec:azimuthal}
        
        \begin{figure}[h!]
                \centering
                \includegraphics[width=9cm]{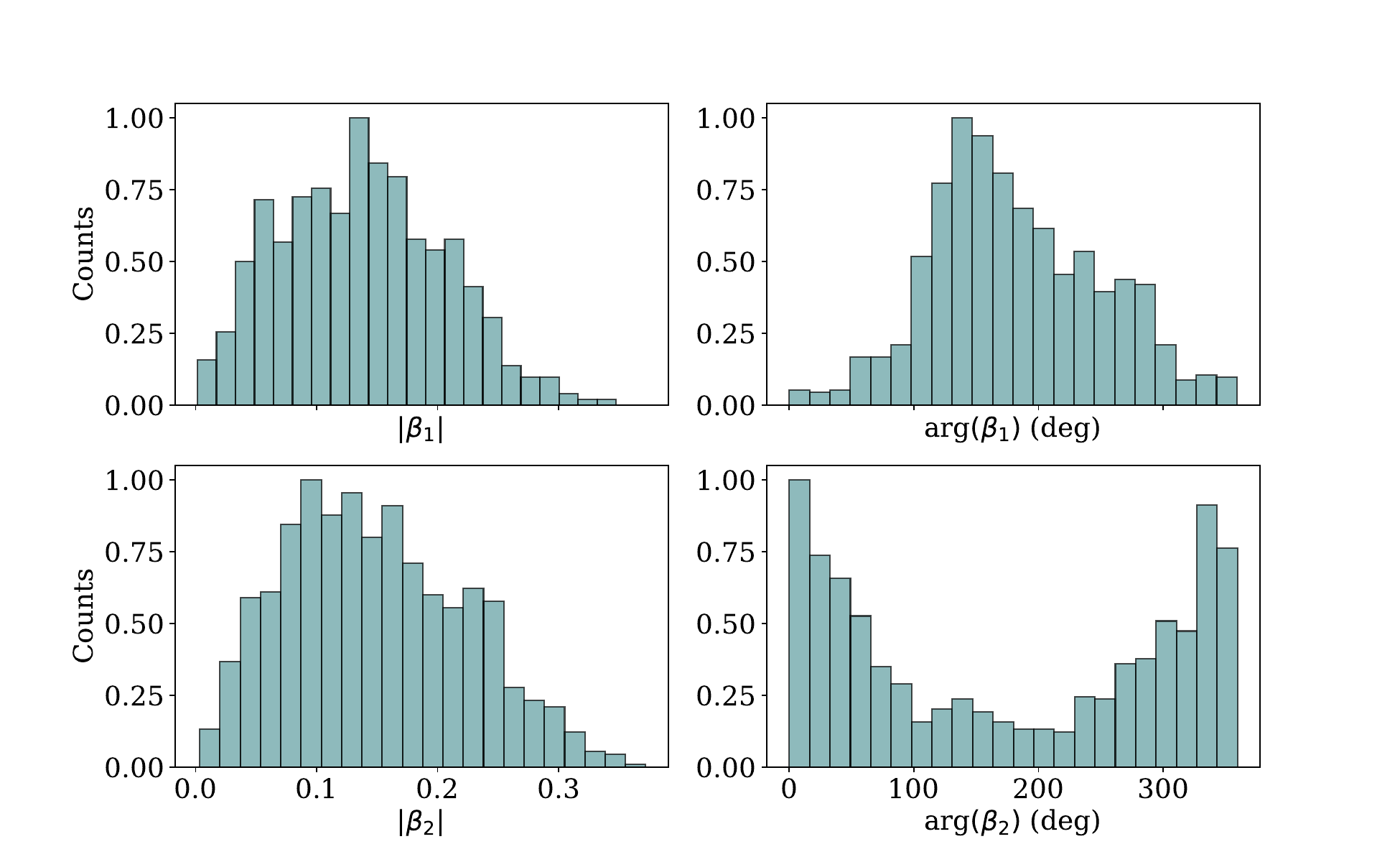}
                \caption{Normalized distribution of the absolute values and argument for the $m=1,2$ beta modes resulting from the model fitting of the 86~GHz data.}
                \label{fig:betahist}
        \end{figure}
        
        Figure~\ref{fig:betahist} shows the distribution of the two beta modes resulting from fitting our model to the entire 86~GHz dataset. Note that due to the symmetry of the quadrupole mode, $\arg(\beta_2)$ is ambiguous by 180$^\circ$.
        
        \begin{figure}[h!]
                \centering
                \includegraphics[width=8cm]{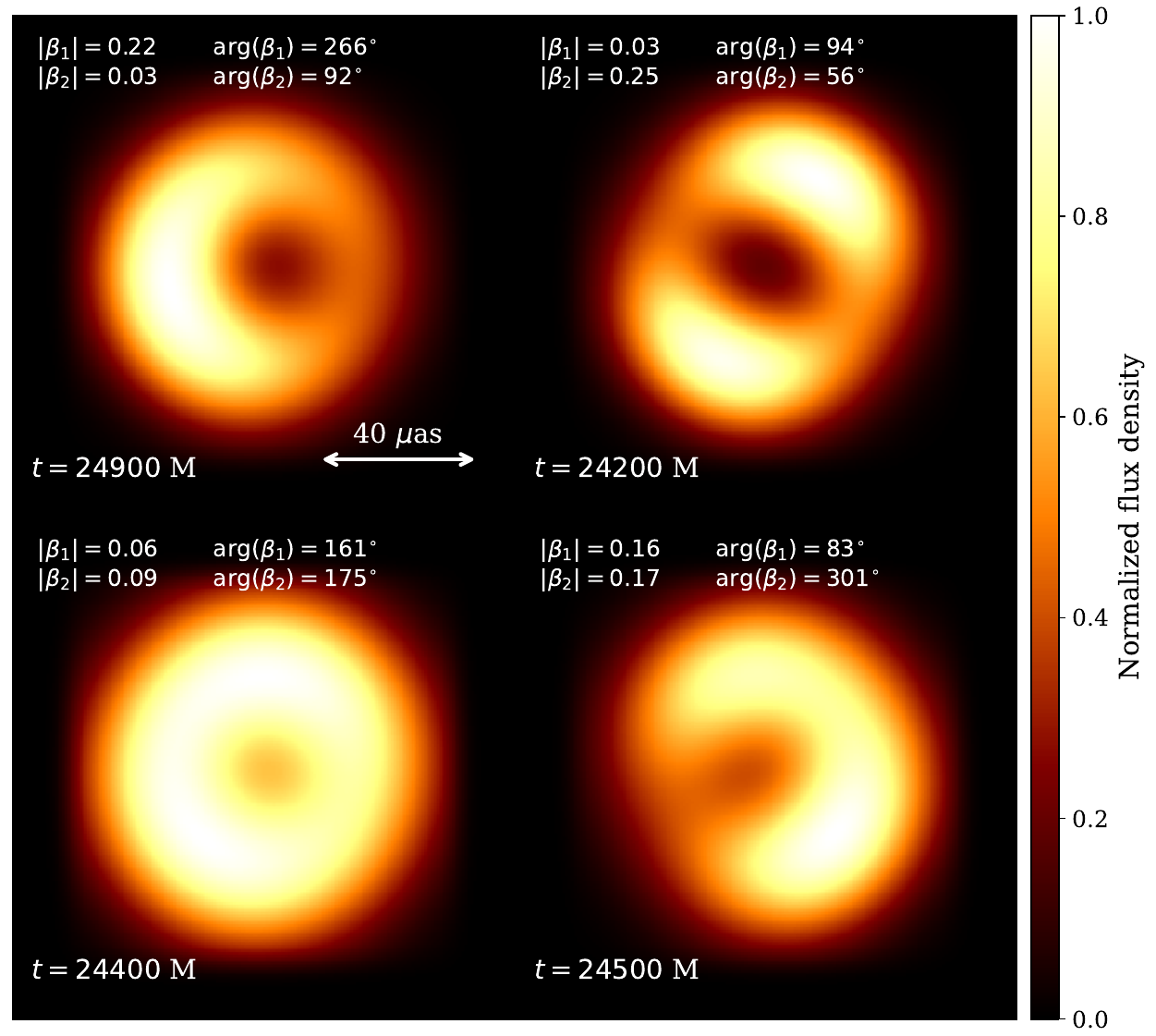}
                \caption{Four time steps that display different ring morphologies (clockwise from the top left): dipole-dominant, quadrupole-dominant, complex, and ring-like. The more complex structure is caused by nonzero beta modes that are comparable in magnitude.}
                \label{fig:azimuthalstructure}
        \end{figure}
        
        The resulting azimuthal structure of a given time step will depend not only on the position of the $m=1$ and $2$ beta modes of Eq.~\ref{eq:mringvis}, corresponding respectively to a dipole and quadrupole mode, but also on their absolute value relative to each other. In a time step where $\left|\beta_1\right|\gg\left|\beta_2\right|$, the structure will take on the shape of a crescent-like structure, with the brightest regions located in the position of $\arg{(\beta_1)}$. Inversely, if  $\left|\beta_1\right|\ll\left|\beta_2\right|$, the emission will be dominated by a quadrupole structure with two hotspots opposite of each other. If both absolute values are nearly zero, the central region will present a nearly symmetrical ring-like shape. Finally, if the two modes have comparable absolute values, the source will present a more complex structure, dependant on the relative position of the angles. This is showcased in Fig.~\ref{fig:azimuthalstructure}. Note that the bottom right image displays a similar structure to that observed by \citet{lu2023}.
        
        While the 230~GHz data follow the same logic, analyzing the physical meaning of the beta modes is less straightforward due to the extra degrees of freedom in our model and the effect of the free Gaussians on the ring morphology.

        \subsection{Role of nonthermal emission and magnetic flux eruption events}
        \label{sec:nonthermal}
        
        \begin{figure}[h!]
                \includegraphics[width=9cm]{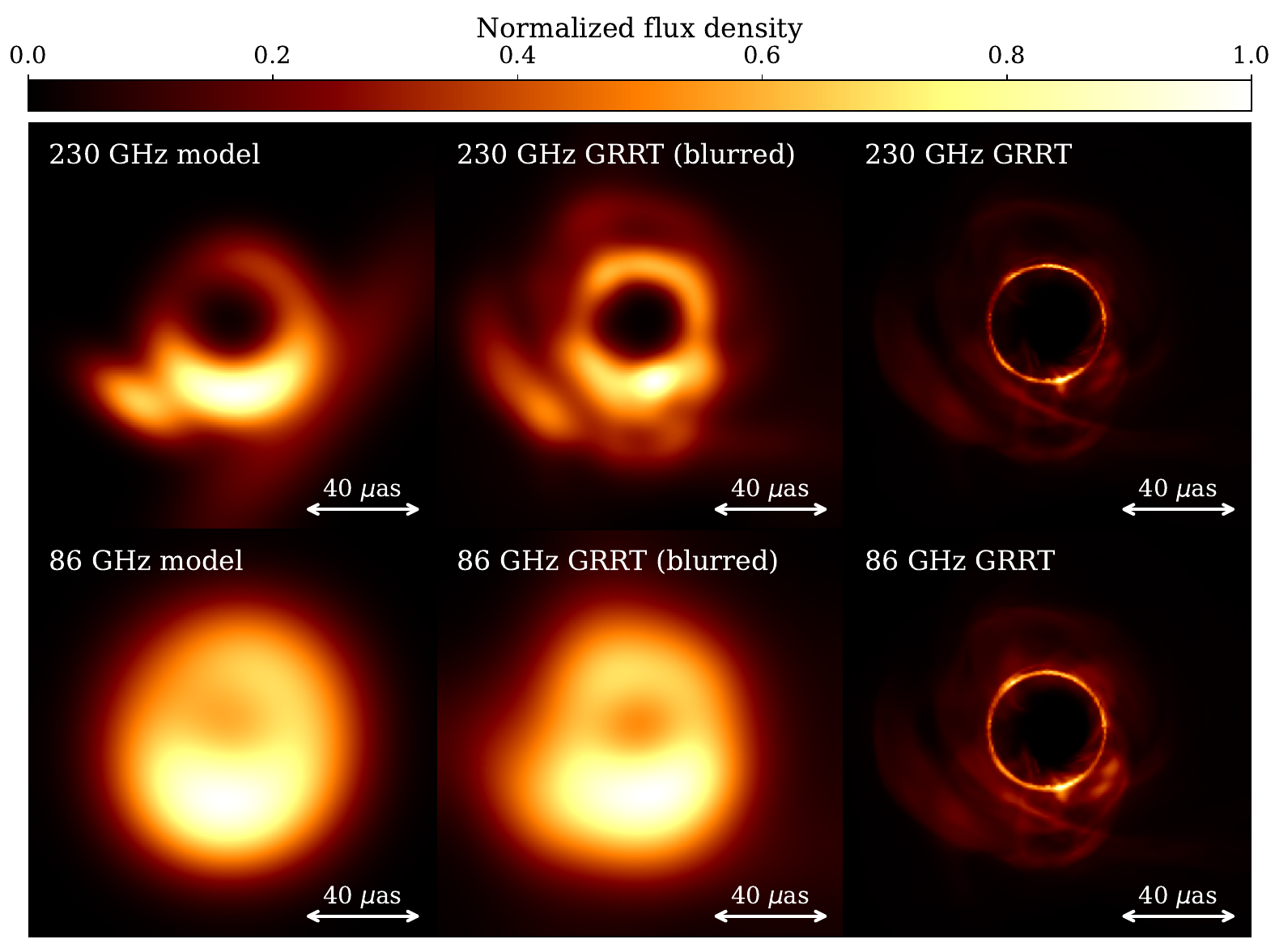}
                \caption{Comparison of the underlying GRRT data and the output of our model fitting at 230\,GHz (top) and 86\,GHz (bottom) for $t=29800\,M$. Right column: Output of the GRRT calculations, non-blurred. Central column: Same GRRT images, blurred by a beam of 10$\,\mu$as and 27$\,\mu$as for 230~GHz and 86~GHz, respectively. Left column: Parametrized model resulting from our fitting procedure, blurred with those same beams. All plots have been normalized to their maximum flux for ease of comparison. Note the increased surrounding flux at 86~GHz in the right column.}
                \label{fig:2980_sixcomparison}
        \end{figure}
        
        Figure~\ref{fig:2980_sixcomparison} shows the 230\,GHz and 86\,GHz image structure at 29800\,M, where  each plot is normalized to its maximum flux. The GRRT images and their corresponding parametrized models have both been blurred with the same beam for each frequency. The increased resolution of 230~GHz observations allows for the surrounding emission to be separated from the underlying ring, while these elements blend together at 86~GHz, resulting in a ring of larger apparent diameter. At the same time, the non-blurred GRRT images show the additional effect of optical depth at these frequencies. The surrounding emission becomes fainter at 230~GHz, while it could be a significant contribution to the overall morphology at 86~GHz.
        
        To investigate these effects, we focused on one of the time steps with a diameter that fell within the observational values, $t=24000\,M$, with a diameter of $60.3\pm0.3\,\mu$as. In the upper limit of $GM/Dc^2=4.2\,\mu$as, this increased to $64.9\pm0.3\,\mu$as. The data at the central $GM/Dc^2$ value, as seen in Fig.~\ref{fig:hist}, contains diameters that fall within the range calculated by \citet{lu2023} for 36\% of the time steps, and for the narrower range calculated with \texttt{resolve} by \citet{Kim2025} for 12\%. For the upper $GM/Dc^2$ limit, seen in Fig.~\ref{fig:highMD}, this respectively becomes 86\% and 43\% of the time steps.
        
\begin{figure*}[h!]
                \includegraphics[width=\textwidth]{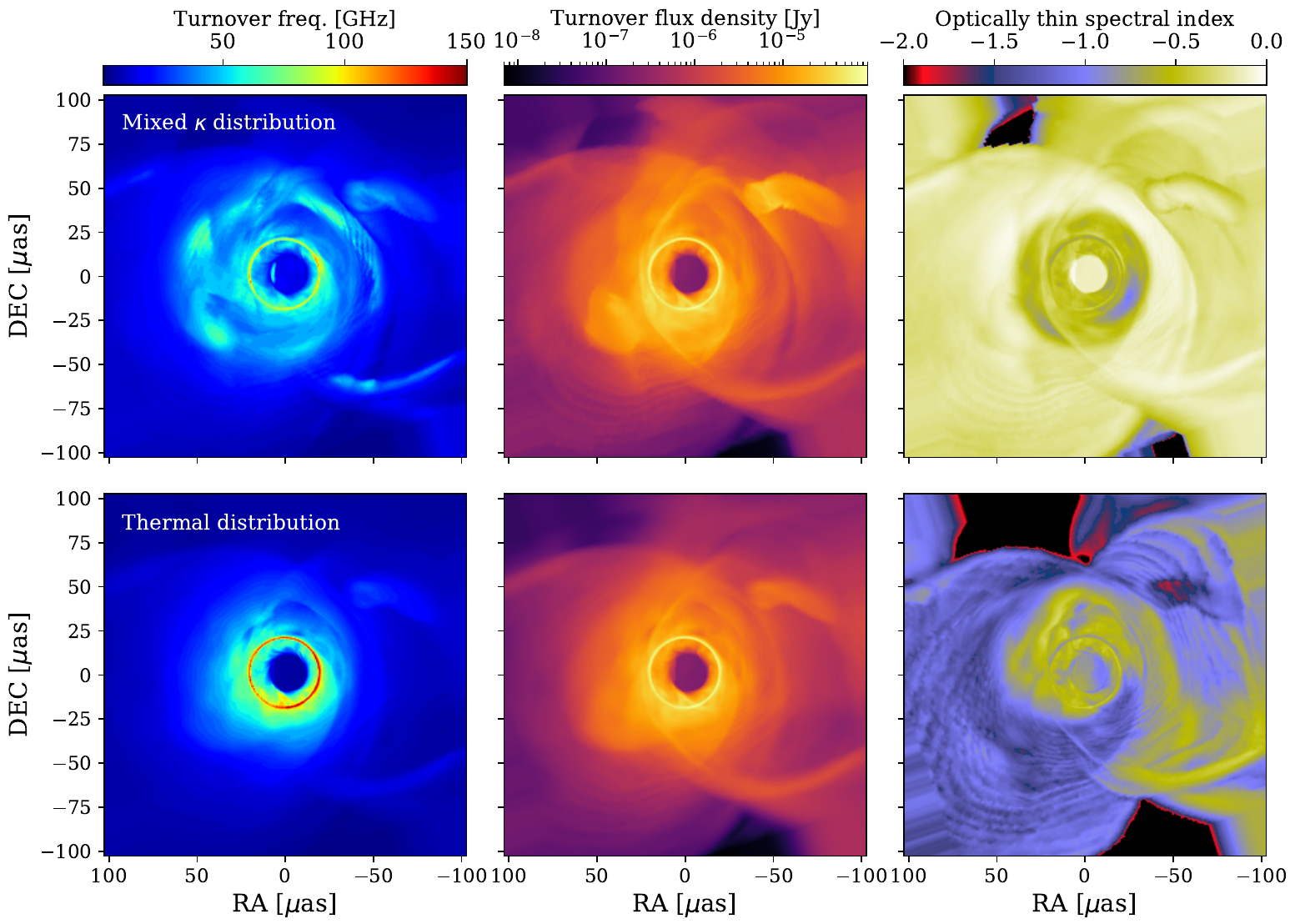}
                \centering
                \caption{Distribution of the turnover frequency, turnover flux density, and optically thin spectral index for a mixed eDF of thermal and kappa distribution (top) and thermal eDF only (bottom) at $t=24000\,M$.}
                \label{fig:turnover}
        \end{figure*}

        To examine the cause of this, Fig.~\ref{fig:turnover} shows spectral information at the central 200$\,\mu$as for $t=24000\,M$. We computed these maps with logarithmically spaced frequencies between 100~MHz and 500~GHz using 100 bins. The leftmost plot shows the turnover frequency at which the emission becomes optically thin. The central plot shows the flux density at this frequency. The rightmost plot shows the spectral index between the turnover frequency and 500 GHz.

        The upper row of Fig.~\ref{fig:turnover} shows our usual GRRT calculations, with a mix of thermal and nonthermal electrons as modeled by a $\kappa$ distribution. The leftmost plot shows turnover frequencies of $\sim\,50-60$~GHz within a radius of $r\sim50\,\mu$as . Most of the flux is distributed roughly within the same radius, as seen in the turnover flux density plot in the middle panel. Interestingly, the optically thin spectral index is steep, $\sim-0.5$, within $r\sim30-40\,\mu$as while a flat value of $\sim 0$ is obtained outside. Note that we injected nonthermal particles at a distance of $r_{\rm inj}=10\,M=38\,\mu$as, which explains the flattening of the spectral index on larger distances due to the power-law tail of the kappa distribution and our choice of the kappa parameter (see Eq. \ref{eq:kappapara}). 
        
        Using this time step as a laboratory, we reran our GRRT calculations on the same GRMHD data, this time only computing the emission and absorption coefficients of a purely thermal electron distribution, that is, assuming the emission to be thermal everywhere. The effects of this on the spectral profile of the $t=24000\,M$ data are seen in the bottom row of Fig.~\ref{fig:turnover}. The turnover frequencies of $\approx60$\,GHz are now concentrated within $r\sim 40\,\mu$as, and are lower outside of that. Changing the eDF has a less apparent impact on the turnover flux density maps, compared to the turnover frequency maps. This is not a surprise, as the kappa-eDF shares the core of the thermal eDF, i.e., most of the emission is still generated by the thermal particles. The most severe differences occur in the optically thin spectral index, especially outside of $r\sim30-40\,\mu$as, where it takes on a value of around $-1.0$, in agreement with the exponential decay of the thermal emissivity $j_{\rm \nu,th}\propto \exp{(-\nu)}$.  These dramatic differences support our conclusion: a combination of a lower turnover frequency in the outer areas of the ring as well as a much steeper spectral index would result in a smaller ring at 86~GHz in the thermal-only case, in contrast to the extended emission structure of the mixed eDF. We ran our thick $m$-ring fitting algorithm to the thermal-only $t=24000\,M$ data and found a ring diameter of $51.7\pm0.4\,\mu$as, a significant decrease from $60.3\pm0.3\,\mu$as, supporting our findings based on the spectral properties. 
        
        \begin{figure*}[t!]
                \includegraphics[width=\textwidth]{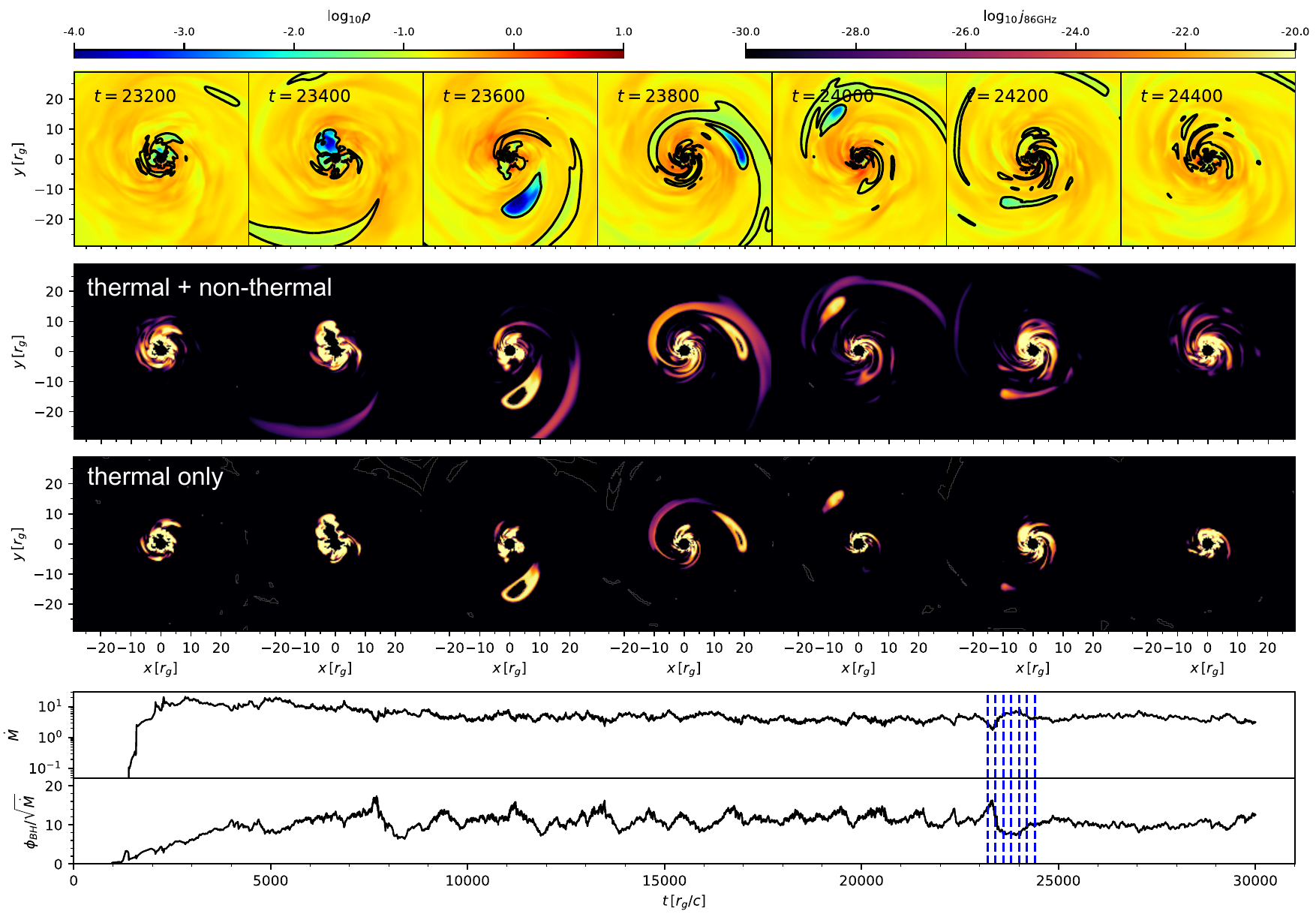}
                \centering
                \caption{Evolution of a MAD event in the equatorial plane. Top row: Logarithm of the density with overlaid contours indicating locations of particle acceleration, where $\tilde{\epsilon}=\epsilon_\textrm{eff}$. Second and third rows: Logarithm of the total emissivity, for a thermal plus kappa eDF and thermal-only eDF using the parameters listed in Table \ref{table:grrtparams}. Two bottom panels: Evolution of the mass accretion rate, $\dot{M,}$ and the MAD parameter, $\phi_{\rm BH}/\sqrt{\dot{M}}$. The vertical blue lines indicate the temporal position of the top panels. }
                \label{fig:MADevent}
        \end{figure*}
        
        However, the question arises of which plasma events cause the extent of the flux in the emission region, how long do they last and how common are they. To examine this, we analyzed the plasma properties between 23200\,M and 24400\,M. In Fig. \ref{fig:MADevent} we present the evolution over several time steps of the plasma and its associated emission in the equatorial planes. MAD simulations are known for frequent ejection events of magnetic flux emerging from a magnetically saturated black hole magnetosphere \citep[see, e.g.,][]{Igumenshchev2008}. These manifest as bundles of poloidal magnetic field penetrating through low density blobs or arcs in the accretion flow of the disk. In our simulations, these blobs are best seen at $t=23600\,M$ (dark blue region). These regions shear out in the accretion flow \citep{Porth2021} and are subject to particle acceleration via magnetic reconnection \citep[see, e.g.,][]{Hakobyan2023}. In addition, the accretion disk is commonly affected by Rayleigh-Taylor instabilities, which are capable of triggering magnetic reconnection at the intersection of the spiral-like accretion streams \citep{Zhdankin2023}. We selected these regions based on the criterion that $\tilde{\epsilon}=\epsilon_\textrm{eff}$ in Eq.~\ref{eq:epsilon}, and plot them as black contours in the top panels of Fig.~\ref{fig:MADevent}. In addition to the flux bundles (blue), we also see the shearing regions in the accretion flows (light green). The corresponding emission coefficients for both our hybrid thermal-nonthermal model and thermal-only model are shown in the two middle rows. Notice that we used the emission model presented in Table \ref{table:grrtparams}. A value of $R_{\rm high}=160$ will lead to a low electron temperature in the disk, resulting in only the innermost regions generating emission, as can be seen in the images. We furthermore applied a magnetization cutoff value of $\sigma_{\rm cut}=5$, rejecting emission from highly magnetized regions. This explains the missing emissivity in the orbiting flux bundles (dark regions within the bundles, best seen at $t=23600\,M$) and the missing emissivity close to the black hole at $t=23400\,M$. We see that the emitting region expands during the flux eruption event, independent from the choice of eDF. After $t\sim800\,M$ ($\sim$280 days for M\,87), the emitting region is back to a size similar to the one before the eruption event. The addition of nonthermal particles enhances the emissivity within the flux bundles and adds additional emissivity into the magnetized accretion streams (best seen at $t=23800\,M$), explaining the increased ring diameter estimated for the nonthermal models compared to the thermal ones. This effect is intensified during the flux eruption events due to the aforementioned processes. 
        
        \begin{figure}[h!]
                \includegraphics[width=9cm]{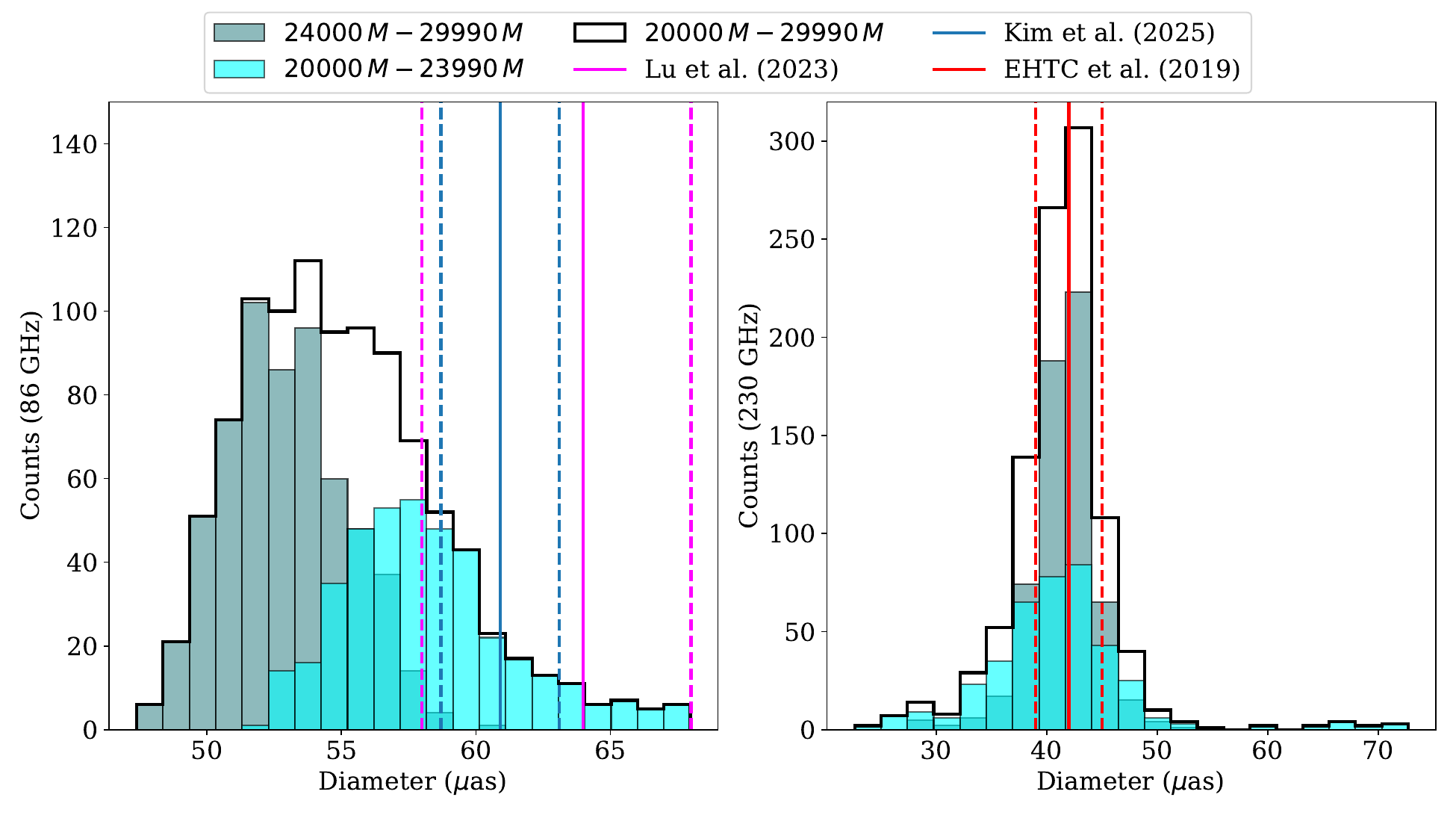}
                \caption{Distribution of the diameters resulting from fitting our models to the 86~GHz and 230~GHz datasets. The data are the same as in Fig.~\ref{fig:hist}, here shown split into the time steps before and after $t=24000\,M$. The black outline shows the total distribution, corresponding to that of Fig.~\ref{fig:hist}. The vertical lines show the observational range for the diameter of the ring, using the same values as Fig.~\ref{fig:hist}.}
                \label{fig:splithist}
        \end{figure}
        
        Examining the evolution of the MAD parameters in Figs.~\ref{fig:grmhd} and \ref{fig:MADevent} shows that our selected time step of $t=24000\,M$ displays a strong depression in the MAD parameter, following a large peak. Before this event, but after the initial stabilization of the accretion, the MAD parameter evolved in peaks and valleys in a quasi-periodical fashion around a stable value. After the large peak at $t=23280\,M$ and following depression, accretion appears perturbed. The average value of the MAD parameter decreases, and so does its variance. The central column of Fig.~\ref{fig:grmhd} and many of the columns in Fig.~\ref{fig:MADevent} show an ``active'' state where MAD bubbles appear, in which electron acceleration takes place resulting in emission further away from the central disk. Meanwhile, the rightmost columns of both plots show a ``quiescent'' state, where these features cannot be found and we expect to retrieve the underlying, smaller ring. To examine this assumption, we split the data shown in Fig.~\ref{fig:hist} before and after this $t=24000\,M$ depression in the MAD parameter, and plotted the results in Fig.~\ref{fig:splithist}. Indeed, while still larger than the 230~GHz diameters, the near entirety of the 86~GHz data after $t=24000\,M$ falls outside of the observational ranges, while 75\% of the data before this time falls inside the range of \citet{lu2023}. Critically, the 230~GHz data do not appear affected by splitting the data in this way: while the 86~GHz data have a median diameter of $57.7^{+2.0}_{-1.8}\,\mu$as before $t=24000\,M$ and $52.8^{+1.6}_{-1.6}\,\mu$as after, 230~GHz yields $41.2^{+2.6}_{-3.7}\,\mu$as and $41.8^{+1.4}_{-1.6}\,\mu$as, respectively. This backs our conclusion that the perceived ring size at 86~GHz is highly sensitive to both the nonthermal particle population and the underlying dynamics of the accretion disk.
        
        \section{Summary and outlook}
        
        In this study we carried out 3D GRMHD jet launching simulations followed by GRRT calculations that incorporated a mixed emission model that took both thermal and nonthermal synchrotron emission into consideration. We studied one thousand time steps over a range of several years to be able to study the apparent change in ring size at 86~GHz in the context of jet launching variability. To provide a more robust analysis of our data, we Fourier-transformed and sampled them with the real observational data of \citet{lu2023}, who studied the source at 86~GHz and found a ring size $\sim$50\% larger than that observed at 230~GHz by the EHT. We fit an azimuthally asymmetric thick $m$-ring model to our full visibility dataset and studied the distribution of the ring parameters, particularly focusing on the diameter. We carried out a similar analysis for our 230~GHz data, though with two free Gaussians added to the thick $m$-ring model to account for surrounding emission. We repeated our calculations to study the upper limits of the mass over distance ratio found by the EHT, $GM/Dc^2=4.2\,\mu$as. We finally studied the variability of the hotspots as well as the spectral profile of the source. Our main findings are as follows:
        
        \begin{enumerate}
                \item With the central value of mass over distance, $GM/Dc^2=3.8\,\mu$as, we find a median ring diameter of $54.6^{+2.7}_{-2.4}\,\mu$as at 86~GHz and $41.6^{+1.8}_{-2.3}\,\mu$as at 230~GHz. While the latter is in agreement with the observational value of $42\pm3\,\mu$as, the former has a small overlap with the observational value of $64^{+4}_{-8}\,\mu$as. 
                \item The angular gravitational radius inferred by the EHT carries an uncertainty of $\sim10\%$, $GM/Dc^2=3.8\pm0.4\,\mu$as. Changing this parameter from $3.8\,\mu$as to $4.2\,\mu$as in our GRRT data and rerunning our fitting procedure yields a median diameter of $60.0^{+2.8}_{-2.4}\mu$as at 86~GHz and $45.4^{+1.9}_{-1.8}\,\mu$as at 230~GHz. The latter is still within the range observed by the EHT, and the former is now in close agreement with the observational values found by \citet{lu2023} as well as, more recently, \citet{Kim2025}.
                \item Strong MAD events can lead to a 86\,GHz ring size in agreement with the observed one. In particular, we studied one of the time steps that matches the observations. Rerunning our GRRT calculations only accounting for thermal emission and examining the resulting spectral profile shows that producing this larger diameter is no longer possible. This indicates that the acceleration of electrons to nonthermal distributions may be necessary to explain the observed morphology of this source.
                \item We focused on the evolution of the MAD parameter over time, as we find that our simulation contains a particularly powerful MAD event that disrupts accretion for an extended amount of time afterward. We find that the inherent variability of a jet launched by a MAD disk greatly impacts the perceived ring size. We specifically find that the size of the 86\,GHz ring in our hybrid eDF model is affected by MAD events and an eruption of magnetic flux arcs, which can significantly increase the ring size. 
                \item Finally, we briefly studied how the variability of both the accreting disk and the jet footpoints impacts the morphology of the ring at 86~GHz. Depending on the state of the source, the central structure could be ring-like, crescent-like, quadrupole-like, or have a more complex structure.
        \end{enumerate}
        
        Future work will focus on improving the nonthermal model, for example by including the effect of electron heating via turbulence or magnetic reconnection. Furthermore, incorporating polarized emission into our GRRT calculations will allow us to further constrain the nature of the 86~GHz ring and to better understand the importance of nonthermal particles, as different eDFs affect emissivities, absorptivities, and rotativities in all polarization channels. We further intend to explore the capabilities of future arrays, such as the ngEHT and the ngVLA, and their role in better understanding the physics of jet-disk coupling.
        
        \begin{acknowledgements}
        This research is supported by the DFG research grant ``Jet physics on horizon scales and beyond" (Grant No.  443220636) within the DFG research unit ``Relativistic Jets in Active Galaxies" (FOR 5195). The numerical simulations and calculations have been performed on \texttt{MISTRAL} at the Chair of Astronomy at the JMU Wuerzburg and on \texttt{iboga} at the Institute for theoretical physics in Frankfurt.
        YM is supported by the National Key R\&D Program of China (grant no. 2023YFE0101200), the National Natural Science Foundation of China (grant no. 12273022), and the Shanghai municipality orientation program of basic research for international scientists (grant no. 22JC1410600). This research is supported in part also by the ERC Advanced Grant “JETSET: Launching, propagation and emission of relativistic jets from binary mergers and across mass scales” (grant No. 884631)
        \end{acknowledgements}
        
        \bibliographystyle{aa} 
        \bibliography{bibliography}
        
        \begin{appendix}
                \section{Emission model}
                \label{sec:emissappendix}
                \begin{figure}[h!]
                        \includegraphics[width=9cm]{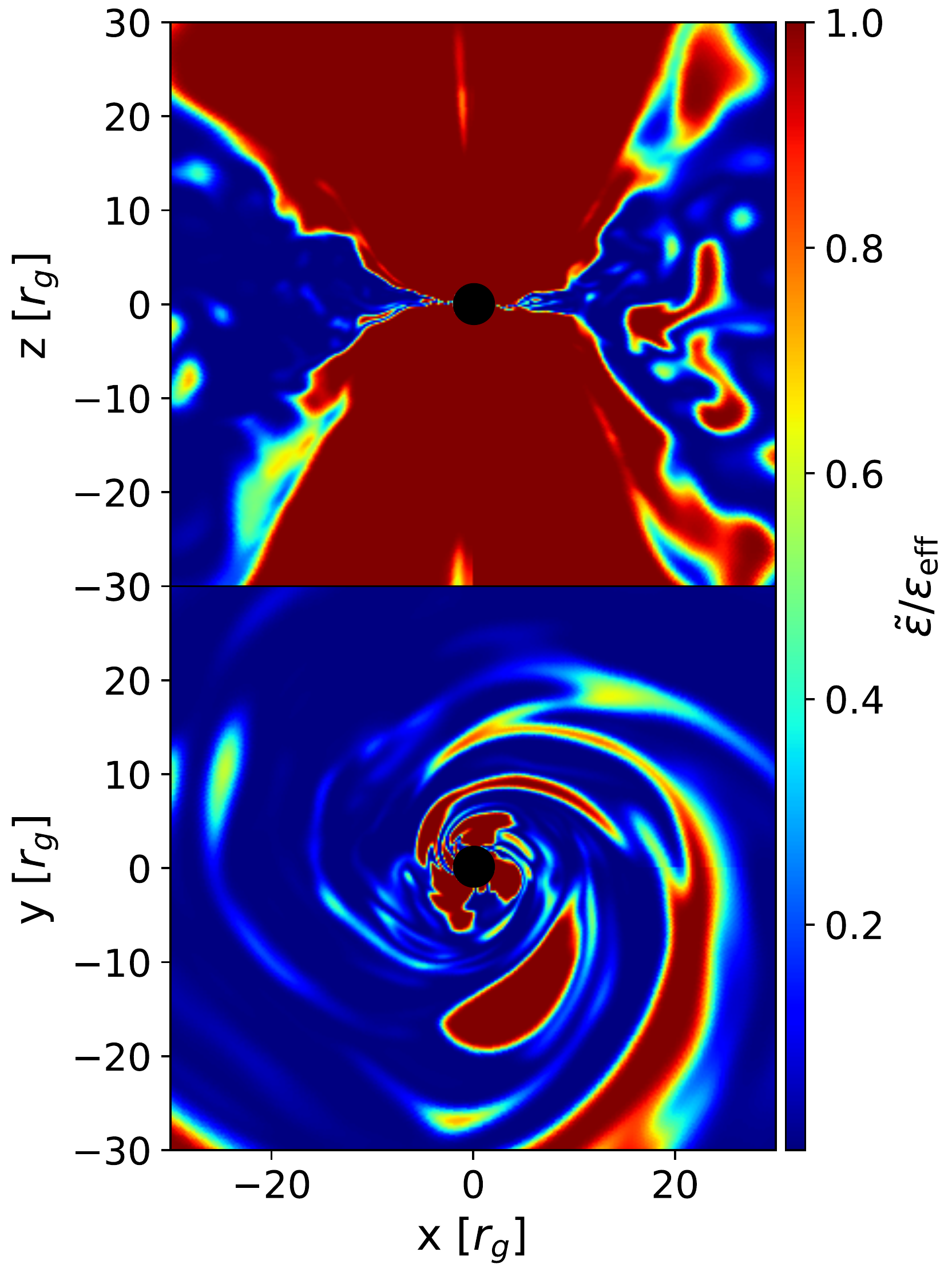}
                        \caption{Meridional (top) and the equatorial (bottom) planes of the central region of the $t=23600\,M$, showing the normalized efficiency ($\tilde{\epsilon}/\epsilon_\textrm{eff}$) as given by Eq.~\ref{eq:epsilon}.}
                        \label{fig:epsiloneff}
                \end{figure}
                
                Figure~\ref{fig:epsiloneff} shows a visual of the partition scheme of our emission model. We computed the parameter $\tilde{\epsilon}$, which governs the partition in Eq.~\ref{eq:emissivity}, and normalized it to its maximum value, $\epsilon_\textrm{eff}$, as seen in Eq.~\ref{eq:epsilon}. It can be seen that, due to its dependence on $\beta$ and $\sigma$, our model largely assigns nonthermal emission to the jet region and thermal emission to the disk region. Ejecta with large values of $\tilde{\epsilon}$ can be seen. $\tilde{\epsilon}/\epsilon_\textrm{eff}=1$ was chosen to plot the black contours in the top panels of Fig.~\ref{fig:MADevent} to highlight regions of particle acceleration.

                \section{Fitting results}
                \label{sec:fittingresults}
                \begin{figure}[h!]
                        \includegraphics[width=9cm]{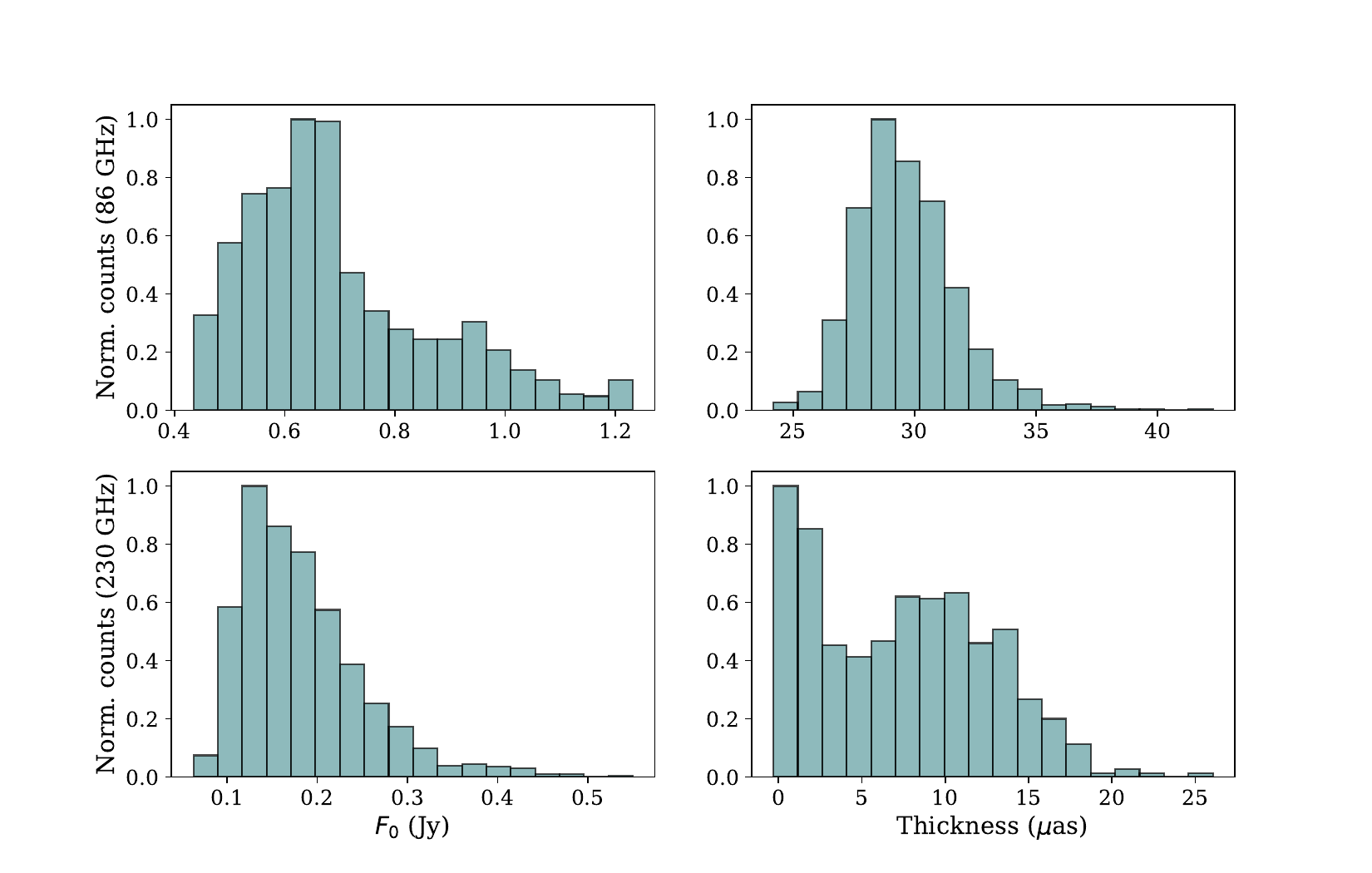}
                        \caption{Normalized distribution of the flux and thickness resulting from fitting our models to the 86~GHz (top) and 230~GHz (bottom) datasets, complementary to Fig.~\ref{fig:hist}.}
                        \label{fig:histfluxalpha}
                \end{figure}
                
                Figure~\ref{fig:histfluxalpha} shows the distribution of values for the ring flux and thickness resulting from fitting our models to both of our datasets, computed at the central value of $GM/Dc^2$. The total flux of the ring is well behaved at 86~GHz and is biased toward lower values at 230~GHz, where the Gaussians carry part of the total flux of the image. Indeed, \citet{EHTf2019} find that the additional Gaussian components in their model can account for anywhere between 10\% and 70\% of the total flux of the 2017 230~GHz data. Similarly, at 86~GHz the ring thickness is normally distributed with a mean value of $29.2\pm1.8\,\mu$as, while the 230~GHz fittings are biased toward thin rings, with half of them returning ring thickness values of $<5\,\mu$as. We hypothesize this to be affected by the Gaussians partly modeling the thickness in the location of the hotspots, while a thinner ring takes care of the underlying structure.
                
                \begin{figure*}[h!]
                        \includegraphics[width=\textwidth]{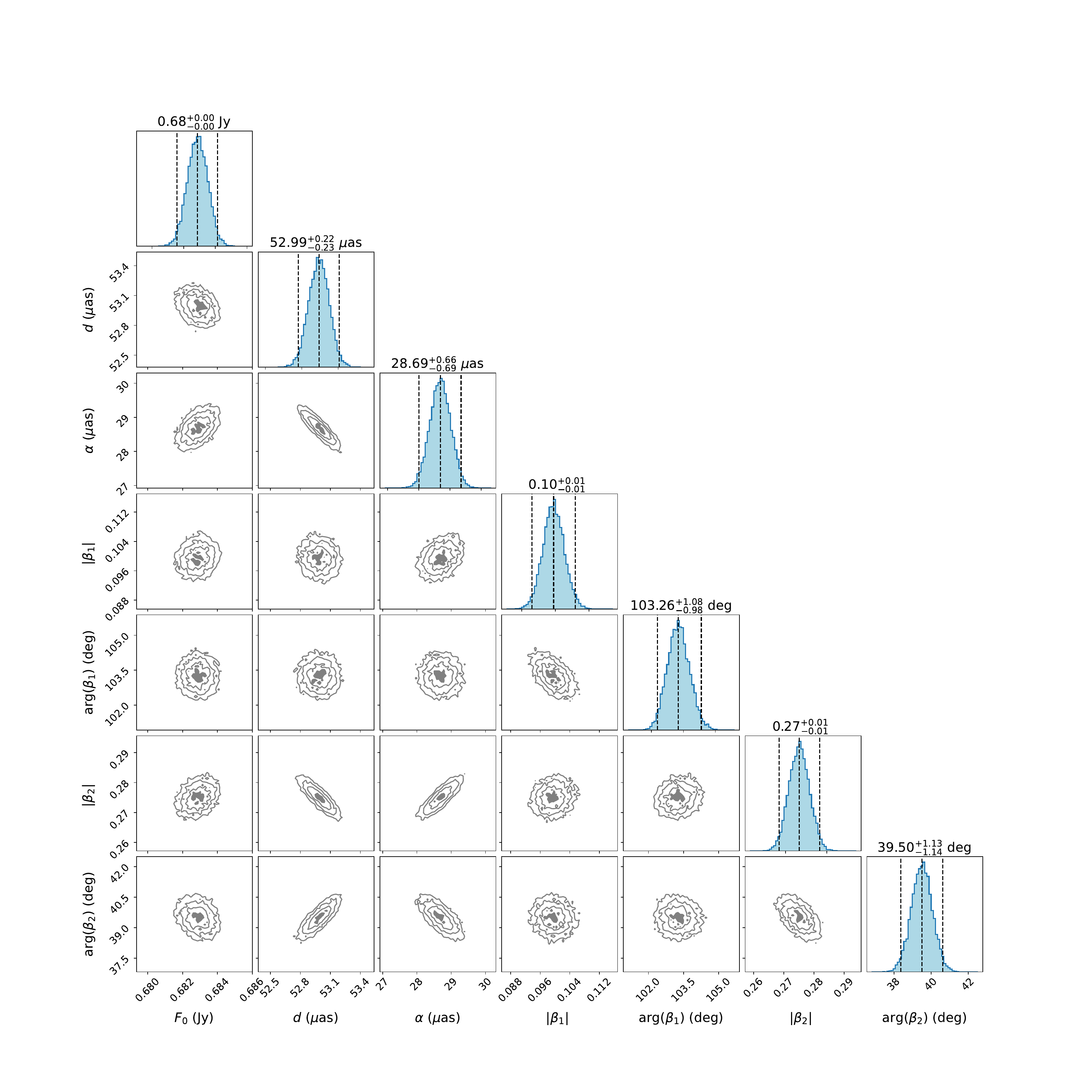}
                        \centering
                        \caption{Bayesian posterior distributions of the thick $m$-ring model to the 86~GHz data at $t=26000\,M$. The diagonal panels show the posterior distributions and the most probable value of each parameter, given as the median value. The vertical dashed lines show this median value as well as the $2\sigma$ range, which is taken as the upper and lower error. The 2D histograms show the joint distributions of each parameter pair (row and column-wise) that shows possible correlations. Contours have been drawn following the samples at different intervals. The innermost shaded area contains samples with a likelihood of $0.5\sigma$, and the outer contours contain the $1\sigma$, $1.5\sigma,$ and $2\sigma$ likelihoods. The total extent of the plots corresponds to a $5\sigma$ likelihood.}
                        \label{fig:cornerplotlarge}
                \end{figure*}
                
                Figure~\ref{fig:cornerplotlarge} shows the full posterior distribution with all seven parameters for a specific time step at 86~GHz. Using \texttt{dynesty} allows us to evaluate the convergence of the model. In this case, the diagonal 1D histograms show that the optimization converged without an issue, returning normal distributions for every parameter. The off-diagonal 2D histograms show correlations between parameters through deviations from a 2D normal distribution. For example, the diameter-thickness plot shows that the Bayesian optimization found an inverse relation between the thickness and the diameter at this time step.
                
                \newpage
                
                \section{Symmetric models}
                \label{sec:symmmodels}
                
                \begin{table}[h]
                        \centering
                        \begin{threeparttable}
                                \caption{Time-averaged $\chi^2$ values for different morphologies and setups on 86~GHz data.}
                                \label{table:chisquareds}
                                \renewcommand{\arraystretch}{1.2}
                                \begin{tabular}{cccc}
                                        \hline \hline \\[-0.9em]
                                        Setup & Morphology & $\chi^2_\mathrm{amp}$ & $\chi^2_\mathrm{cp}$ \\ 
                                        \\[-0.9em] \hline \\[-0.8em]
                                        \multirow{4}{*}{\parbox{2.8cm}{\centering No additional error budget}} & Disk & 57.48 & 46.76 \\ 
                                        & Thin ring & 69.81 & 46.72 \\ 
                                        & Thick ring & 41.96 & 46.60 \\ 
                                        & Thick $m$-ring & 2.55 & 2.90 \\ 
                                        \\[-0.9em] \hline \\[-0.8em]
                                        \multirow{4}{*}{\parbox{2.8cm}{\centering 10\% error budget\\for ALMA baselines}} & Disk & 3.87 & 7.47 \\ 
                                        & Thin ring & 5.20 & 7.83 \\ 
                                        & Thick ring & 3.12 & 6.91 \\ 
                                        & Thick $m$-ring & 0.99 & 0.81 \\ 
                                        \\[-0.9em] \hline
                                \end{tabular}
                        \end{threeparttable}
                \end{table}
                
                We selected the 60 time steps between $t= 24000\,M$ and $t= 30000\,M$, with a cadence of $t_{\rm cad}=100\,M$, and computed the  $\chi^2$ of fitting different models to their synthetic visibilities. Table~\ref{table:chisquareds} shows the $\chi^2$ values of different models, averaged over all sixty time steps. We show the results of fitting to data with the observational errors of \citet{lu2023}, as well as with an additional fractional error of 10\% to the visibilities observed by ALMA.

                \begin{figure}[h!]
                        \includegraphics[width=9cm]{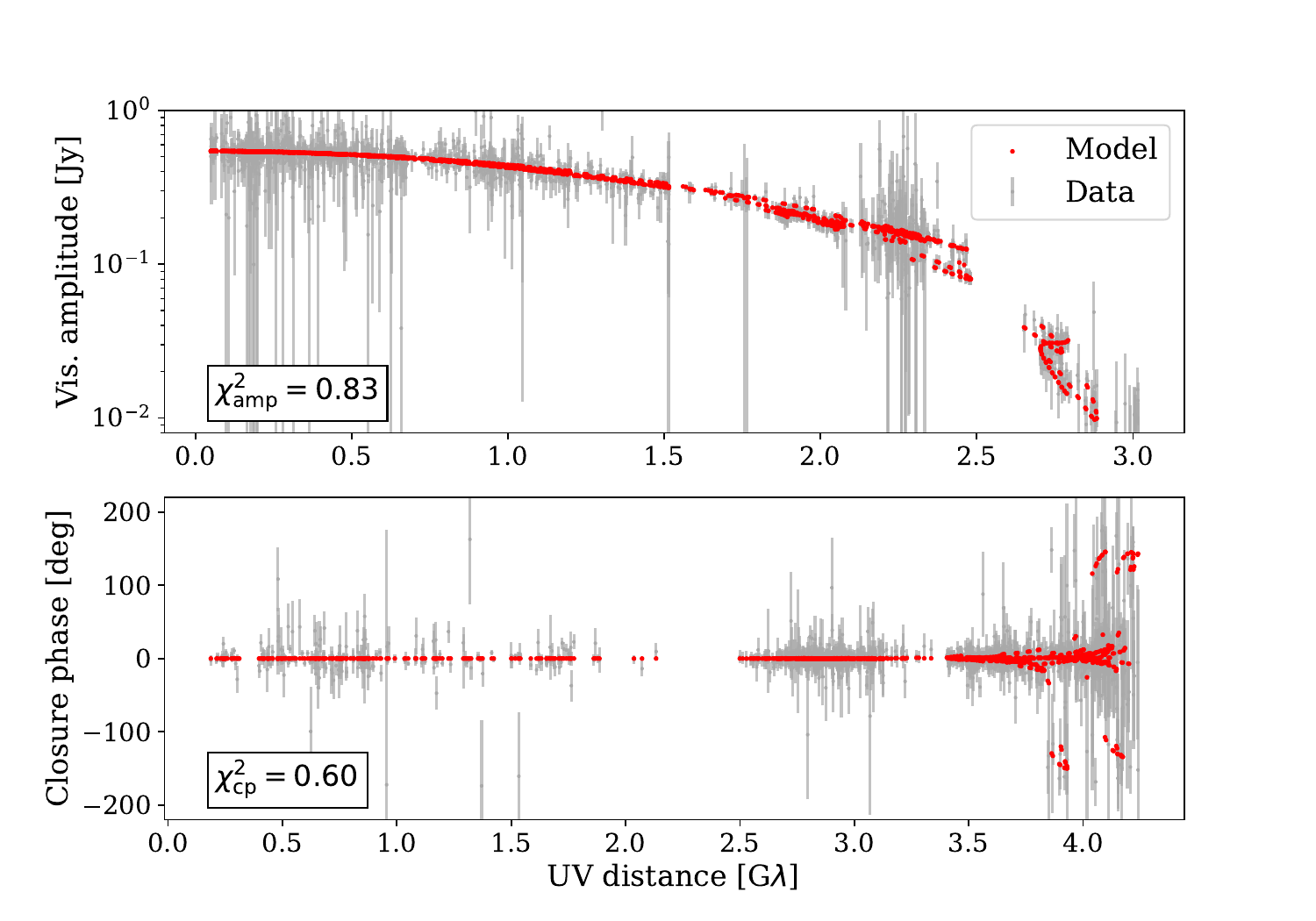}
                        \caption{Visibility amplitude over uv distance for the time step $t=26700\,M$. Overlaid in red is the result of fitting a thick $m$-ring model to the data. Note the relatively few nonzero closure phases.}
                        \label{fig:mringvis2670}
                \end{figure}
                
                \begin{figure}[h!]
                        \includegraphics[width=9cm]{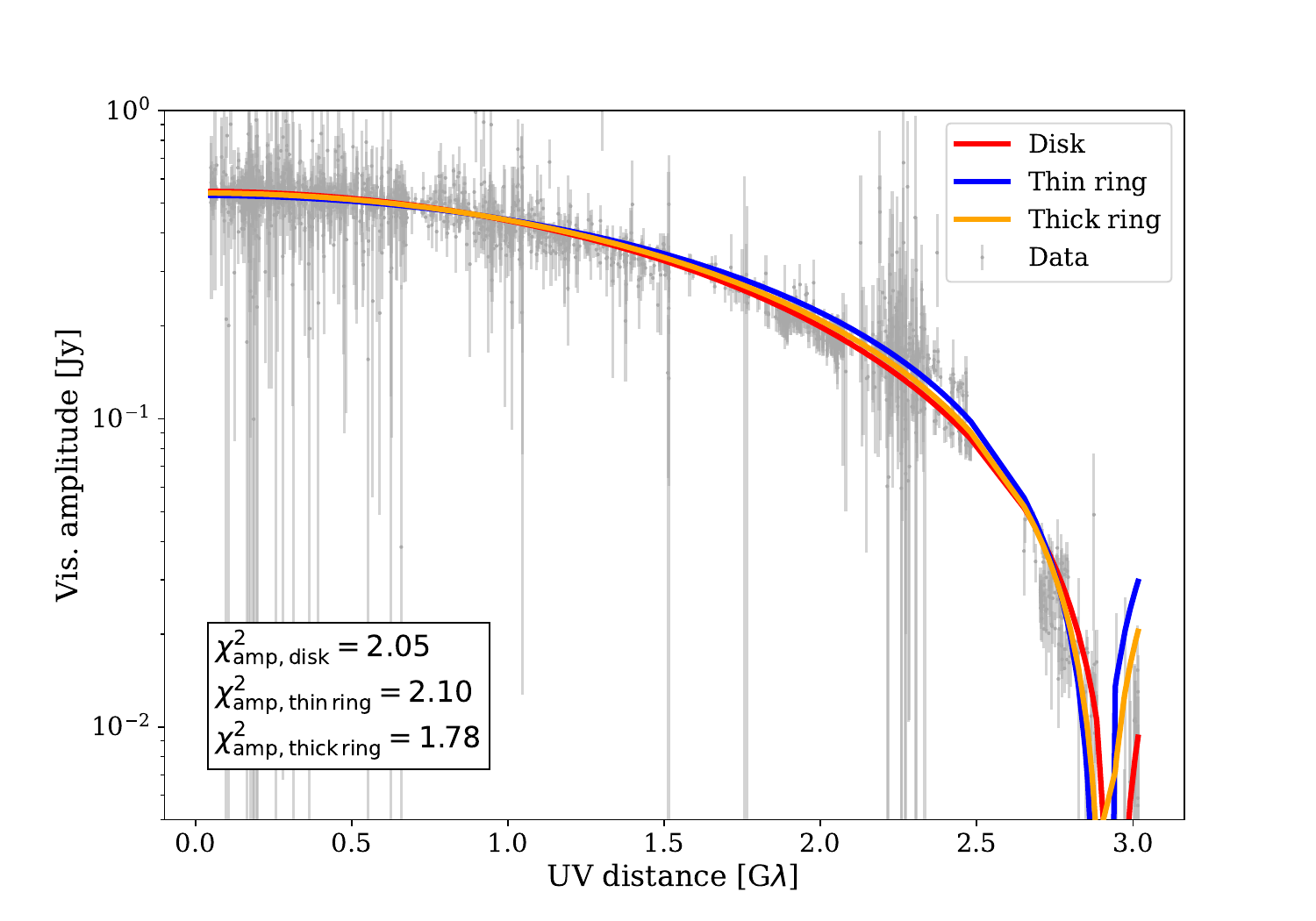}
                        \caption{Visibility amplitudes over uv distance for the image at $t=26700\,M$. Plotted on top are the results of fitting three different azimuthally symmetric models to the data. The $\chi^2_\mathrm{amp}$ values have been computed accounting for an additional 10\% error on ALMA data.}
                        \label{fig:symmfits2670}
                \end{figure}
                
                In addition to the asymmetric thick $m$-ring model, we also fit a top-hat disk, an infinitesimally thin ring, and a thick ring to our test data. Table~\ref{table:chisquareds} shows that adding a 10\% error on ALMA baselines reduces the $\chi^2$ values by an order of magnitude for the symmetric morphologies. While $\chi^2$ are expected to improve with increasing errors, as shown by Eqs.~\ref{eq:chisqamp} and \ref{eq:chisqcp}, it is noteworthy that the thick $m$-ring provides a better fit without additional errors than the symmetric morphologies do with these errors. However, examining the individual time steps shows a large range of $\chi^2$ values. Within our subset, $t=26700\,M$ contains the data with the best $\chi^2_\mathrm{cp}$ values for the symmetric models. With the added ALMA errors, the disk, thin ring and thick ring models obtain $\chi^2_\mathrm{cp}$ values of 0.66, 0.85 and 0.66, respectively. Without the added ALMA errors it is instead $\chi^2_\mathrm{cp}=1.82$, 6.69 and 1.82. The symmetric models also perform relatively well on the amplitudes, with $\chi^2_\mathrm{amp}$ values of 2.05, 2.10 and 1.78 for the disk, thin ring and thick ring, respectively, when we include additional ALMA errors. Without them, we obtain  $\chi^2_\mathrm{amp}$ values of 6.73, 11.35 and 6.66, indicating a bad fit but better than the average of Table~\ref{table:chisquareds}. Figure~\ref{fig:mringvis2670} shows the thick $m$-ring fitting for this time step. Compared to Fig.~\ref{fig:mringvis}, it is clear that there are fewer nonzero closure phases, as expected of a time step where symmetric models would perform well. Indeed, Fig.~\ref{fig:symmfits2670} shows the fit of all three of these models on top of the $t=26700\,M$ amplitudes. The models largely retrieve the features of this time step.
                
                \begin{figure}[h!]
                        \includegraphics[width=9cm]{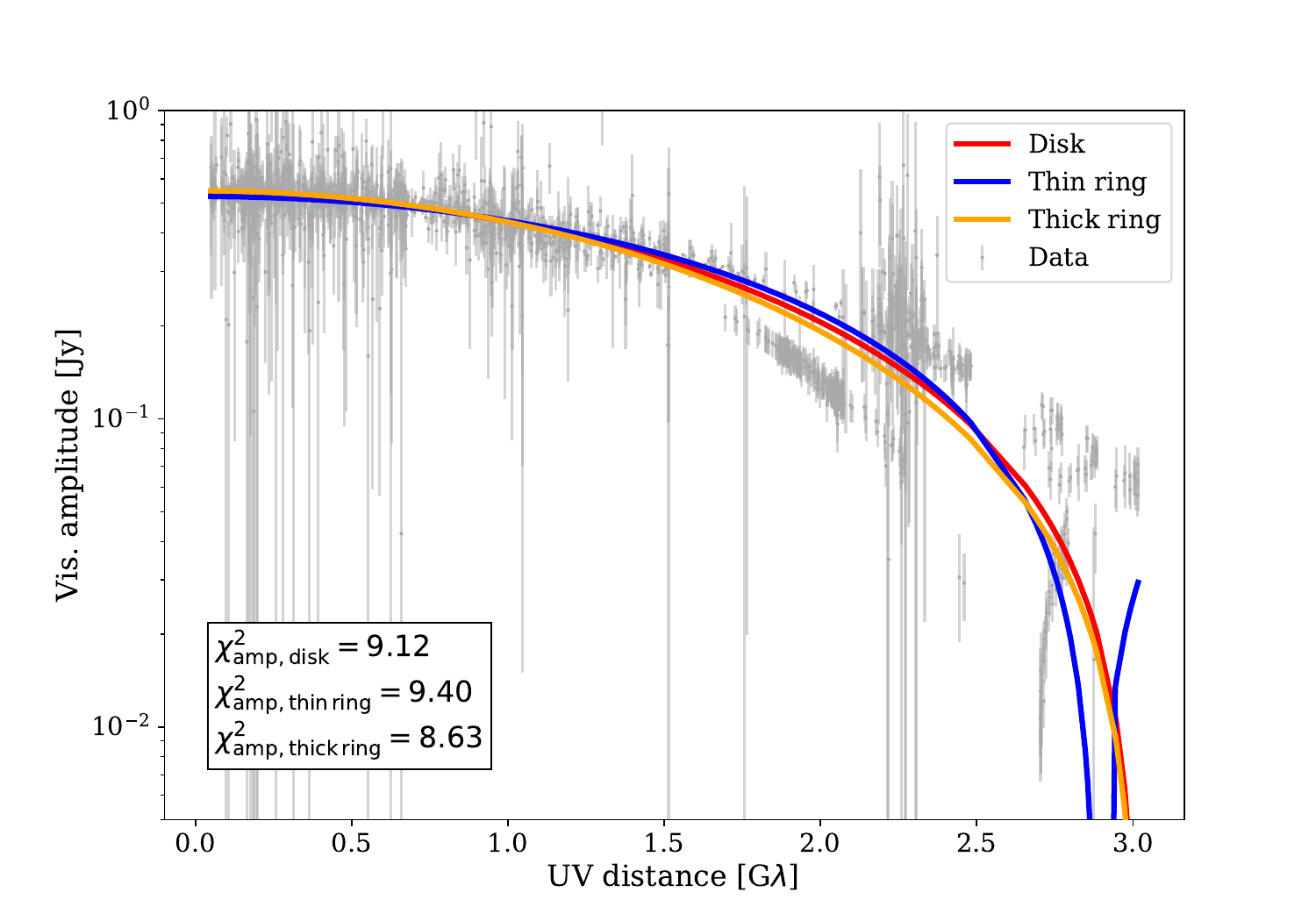}
                        \caption{Visibility amplitudes over uv distance for the image at $t=28600\,M$. Plotted on top are the results of fitting three different azimuthally symmetric models to the data. The $\chi^2_\mathrm{amp}$ values have been computed accounting for an additional 10\% error on ALMA data.}
                        \label{fig:symmfits2860}
                \end{figure}
                
                By comparison, Fig.~\ref{fig:symmfits2860} contrasts the performance of the symmetric morphologies on the same time step as Fig.~\ref{fig:mringvis}, where several asymmetric features are present. Even with ALMA errors, these models return $\chi^2_\mathrm{amp}$ values of around $8-9$. Without errors, the disk, thing ring and thick ring models respectively return $\chi^2_\mathrm{amp}$ values of 189.04, 197.87 and 165.39. The $\chi^2_\mathrm{cp}$ values are all around 26. The data at $t=26700\,M$ have a clear lack of asymmetric features compared to the data at $t=28600\,M$. This asymmetry is visible not only in the nonzero closure phases, but also in structures present in the visibility amplitudes that cannot be fit by unmodulated Bessel functions.
                
                \section{230 GHz model}
                \label{sec:230ghzmodel}
                
                We used the same test dataset at 230~GHz as the one used at 86~GHz in Appendix~\ref{sec:symmmodels}. Applying the simple thick $m$-ring model to 230~GHz data yielded average $\chi^2$ values of $\chi^2_\mathrm{amp}=37.4$ and $\chi^2_\mathrm{cp}=23.0$. The inclusion of two free Gaussian components brings our fit to $\chi^2_\mathrm{amp}=1.71$ and $\chi^2_\mathrm{cp}=1.85$.
                
                \begin{figure}[h!]
                        \includegraphics[width=9cm]{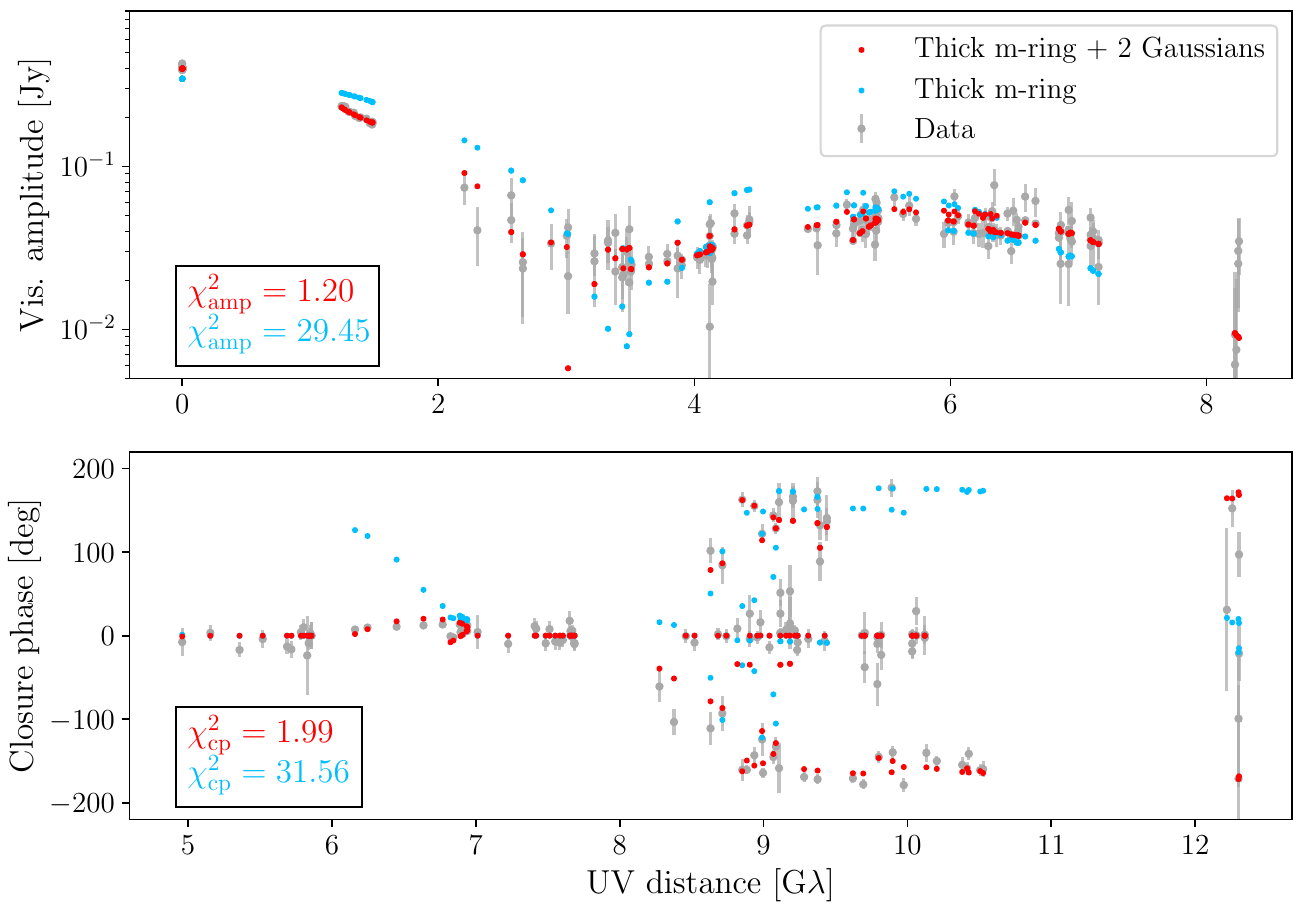}
                        \caption{Visibility amplitudes and closure phases over uv distance for the time step $t=28900\,M$. Overlaid in blue is the result of fitting the same $m=2$ thick $m$-ring that we previously fit to 86~GHz data. The model corresponding to the red data points, by contrast, contains two free Gaussian features.}
                        \label{fig:visibilities230GHz2890both}
                \end{figure}
                
                Figure~\ref{fig:visibilities230GHz2890both} compares the result of fitting an $m=2$ thick $m$-ring model to the same time step with and without the free Gaussians. The extra Gaussian features are particularly necessary for modeling the visibility amplitudes between 1.2 and 1.5~G$\lambda$, corresponding to the LMT-SMT (Submillimeter Telescope) baseline, as well as the closure phases between 6 and 7~G$\lambda$, corresponding to the ALMA-SMT-LMT triangle.
                
                \section{Impact of the field of view}
                \label{sec:fieldofview}
                
                \begin{figure}[h!]
                        \includegraphics[width=9cm]{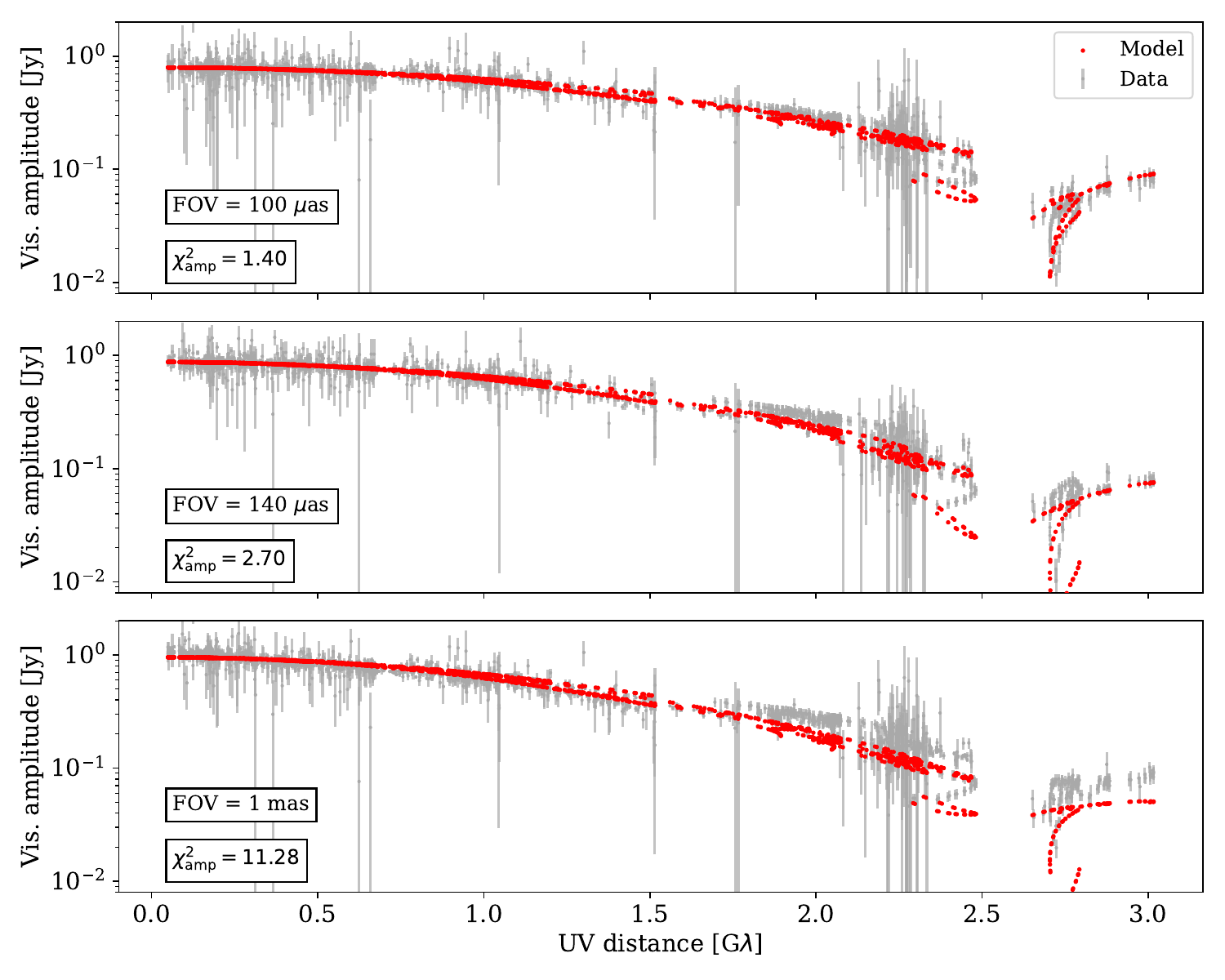}
                        \caption{Visibility amplitudes over uv distance for the time step $t=24500\,M$. Overlaid in red is the thick $m$-ring model fit to each set of visibilities. Each panel corresponds to visibilities computed from the central image with different fields of view.}
                        \label{fig:fov86}
                \end{figure}
                
                Figure~\ref{fig:fov86} shows the results of fitting our $m$-ring model to the visibilities of a representative time step, computed from the central 100$\,\mu$as, 140$\,\mu$as and 1~mas. Increasing the FOV incorporates more of the bulk emission into the visibilities, resulting in higher amplitudes at the shortest baselines and more complex, less ring-like structures at the longest baselines. The fit to the longest baselines is critical to study the diameter of a ring model, as this quantity is related to the first nulls in the Bessel functions that comprise the Fourier transform of a ring. Similarly, the second nulls are related to the thickness of said ring. As this uv coverage does not contain long enough baselines to probe these second nulls, the curvature of the data after the first null at the longest baselines is all that the model has to estimate the thickness of the ring. We find that increasing the FOV not only worsens the quality of the fit, but it also increases the thickness while keeping the diameter roughly constant. On our test dataset, performing the fitting on data with FOV = 100$\,\mu$as, 140$\,\mu$as, and 1~mas resulted in median diameters of 53.1~$\mu$as, 53.2$\,\mu$as, and 55.6~$\mu$as, respectively. On the other hand, the fits returned for the median values of the thickness 28.8$\,\mu$as, 38.6$\,\mu$as, and 46.3$\,\mu$as. This shows a high sensitivity of the ring thickness on the selected FOV, which is expected due to the limited uv coverage at the necessary spatial frequencies. We decided on 100$\,\mu$as based on the quality of the fits so as to prevent the bulk of the extended emission from dominating the fit.

        \end{appendix}
\end{document}